\documentclass[journal=jpclcd,manuscript=article]{achemso}
\usepackage[utf8]{inputenc}
\usepackage[T1]{fontenc}
\usepackage[version=3]{mhchem}
\usepackage{soul}
\usepackage{bm}
\usepackage{threeparttable}
\usepackage{xcolor}

\makeatletter
\begin{tocentry}
\includegraphics[width=2in]{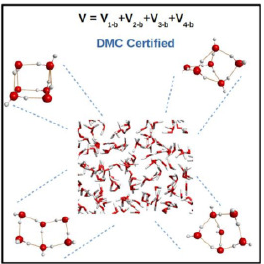}
\end{tocentry}

\author{Qi Yu}
\affiliation{Department of Chemistry Yale University, New Haven, Connecticut 06520, U.S.A.}
\email{q.yu@yale.edu}
\author{Chen Qu}
\affiliation{Independent researcher, Toronto, Ontario, Canada}
\email{szquchen@gmail.com}
\author{Paul L. Houston}
\email{plh2@cornell.edu}
\affiliation{Department of Chemistry and Chemical Biology, Cornell University, Ithaca, New York
14853, U.S.A. and Department of Chemistry and Biochemistry, Georgia Institute of
Technology, Atlanta, Georgia 30332, U.S.A}
\author{Riccardo Conte}
\email{riccardo.conte1@unimi.it}
\affiliation{Dipartimento di Chimica, Universit\`{a} degli Studi di Milano, via Golgi 19, 20133 Milano, Italy}
\author{Apurba Nandi}
\affiliation{Department of Chemistry and Cherry L. Emerson Center for Scientific Computation, Emory University, Atlanta, Georgia 30322, U.S.A.}
\author{Joel M. Bowman}
\email{jmbowma@emory.edu}
\affiliation{Department of Chemistry and Cherry L. Emerson Center for Scientific Computation, Emory University, Atlanta, Georgia 30322, U.S.A.}

\title{q-AQUA: a many-body CCSD(T) water potential, including 4-body interactions, demonstrates the quantum nature of water from clusters to the liquid phase
}

\begin{document}

\clearpage


\begin{abstract}Many model potential energy surfaces (PESs)  have been reported for water; however, none are strictly from ``first principles''.  Here we report such a potential, based on a many-body representation at the CCSD(T) level of theory up to the ultimate 4-body interaction. The new PES is benchmarked for the isomers of the water hexamer for dissociation energies, harmonic frequencies, and unrestricted diffusion Monte Carlo (DMC) calculations of the zero-point energies of the Prism, Book and Cage isomers. Dissociation energies of several isomers of the 20-mer agree well with recent benchmark energies.  Exploratory DMC calculations on this cluster verify the robustness of the new PES for quantum simulations.  The accuracy and speed of the new PES is demonstrated for standard condensed phase properties, i.e., the radial distribution function and the self-diffusion constant.  Quantum effects are shown to be substantial for these observables and also needed to bring theory into excellent agreement with experiment. 
\end{abstract}

\flushbottom
\maketitle

\thispagestyle{empty}
\clearpage
Potential energy surfaces (PESs) are critical to our understanding of molecular interactions, their dynamics, and their structures. Among these surfaces, perhaps the most important are those that predict the behavior of life's signature molecule, water. Ideally such a PES would employ the highest level of electronic structure theory and be developed for a complete many-body interaction.  A major step in this direction was the CC-pol potential published in 2007.\cite{Bukowski2007}  This potential was based on fits to CCSD(T) interaction energies for rigid monomers at the 2-body(b) level of interaction and sophisticated treatments of long-range many-body induction effects.  Since then numerous studies have examined the importance of 3-b, 4-b, and 5-b interactions, and the latest work\cite{MBE20} shows definitively that truncating at the 4-body level accounts for virtually all of the many-body interactions.  So, it is clear now that an ideal approach for a water PES would include 2-b, 3-b and 4-b interactions for flexible monomers and using CCSD(T) level of theory. 
In addition, studies of structural and transport properties of water, ranging from clusters to condensed phase, should ideally be based on quantum simulations, which require a robust PES reaching to energies well beyond the zero-point energy. 
Finally, the PES should be invariant with respect to permutation of monomers, and each monomer should also be invariant with respect to interchange of the two H atoms. 

We report such a PES here and apply it to a variety of important ``stress tests'' for clusters and the condensed phase. The form of this PES is based on the well-known, many-body expression for the total energy of $N$ water monomers:
\begin{equation}
  V(1,\cdots,N)=\sum_{i=1}^NV_{1-b}(i)+\sum_{i>j}^NV_{2-b}(i,j)+\sum_{i>j>k}^NV_{3-b}(i,j,k)+\sum_{i>j>k>l}^NV_{4-b}(i,j,k,l) + \cdots,
\end{equation}
We indicate terms up to 4-b interactions explicitly because the potential in the present paper is  truncated at this term.  Each of these terms is obtained using a machine-learned fit to corresponding datasets of CCSD(T) interaction energies. Specifically, the fits are done using a basis of permutationally invariant polynomials (PIPs).\cite{Braams2009, PIPSJCP22} This particular ML method has been used by us previously in developing a water potential, denoted WHBB,\cite{WHBB} and by Paesani and co-workers for the MB-pol water potential.\cite{mbpol2b,mbpol3b}  These potentials are truncated at the \textit{ab initio} level of 3-b interactions. Also, both use semi-empirical TTMn-F water potentials for higher-body interactions,  however, in different ways.  The WHBB PES uses TTM3-F\cite{TTM3F} while the MB-pol PES uses TTM4-F.\cite{TTM4}  There are significant differences in how these potentials are used in WHBB and MB-pol and these are described in detail in the Supporting Information (SI). Finally we note that there are numerous empirical water potentials, and we refer the reader to a recent review\cite{mbreview} of these along with the WHBB and MB-pol potentials.

The new fits reported here make use of new 2-b and 3-b datasets, which are at the CCSD(T) level and more extensive both in energy and range than the CCSD(T) 2- and 3-b datasets used for the MB-pol potential. In addition, a 4-b dataset, employed by us in a preliminary CCSD(T) PIP 4-b PES,\cite{4b21} is extended and a new 4-b PIP fit is reported. Here we note the numbers of CCSD(T) energies in the datasets are 71,892, 45,332 and 3692 for the 2-, 3- and 4-b interactions, respectively. Additional details are given in the last section and in the SI. These new 2-, 3-, and 4-b PIP PESs together with the spectroscopically accurate Partridge and Schwenke\cite{PS} water monomer (1-b) PES constitute the new PES.  The new potential is denoted ``q-AQUA''. 


We now demonstrate the accuracy and robustness of q-AQUA for standard ``stress'' tests, namely the dissociation energies, harmonic frequencies, and diffusion Monte Carlo (DMC) calculations of zero-point energies of isomers of the water hexamer and the dissociation energies of several isomers of the 20-mer, for which benchmark values have recently been reported.\cite{heindeljpcl}  We also report molecular dynamics (MD) and path integral molecular dynamics (PIMD) calculations for the radial distribution function (RDF) and MD and semi-quantum Ring Polymer MD (RPMD) calculations of the diffusion constant over a range of temperatures. Significant 4-b and quantum effects are found for these properties. 


MB-pol is a highly successful water potential and so we present selected results using that potential as part of the assessment of q-AQUA.  The first  comparisons are for the 2-b interaction with CCSD(T) benchmark calculations. For the 2-b interaction, attractive and repulsive cuts are presented in the SI where both q-AQUA and MB-pol are shown to be in excellent agreement with the CCSD(T) calculations. 

\begin{figure}[htbp!]
\begin{center}
\includegraphics[width=1.0\textwidth]{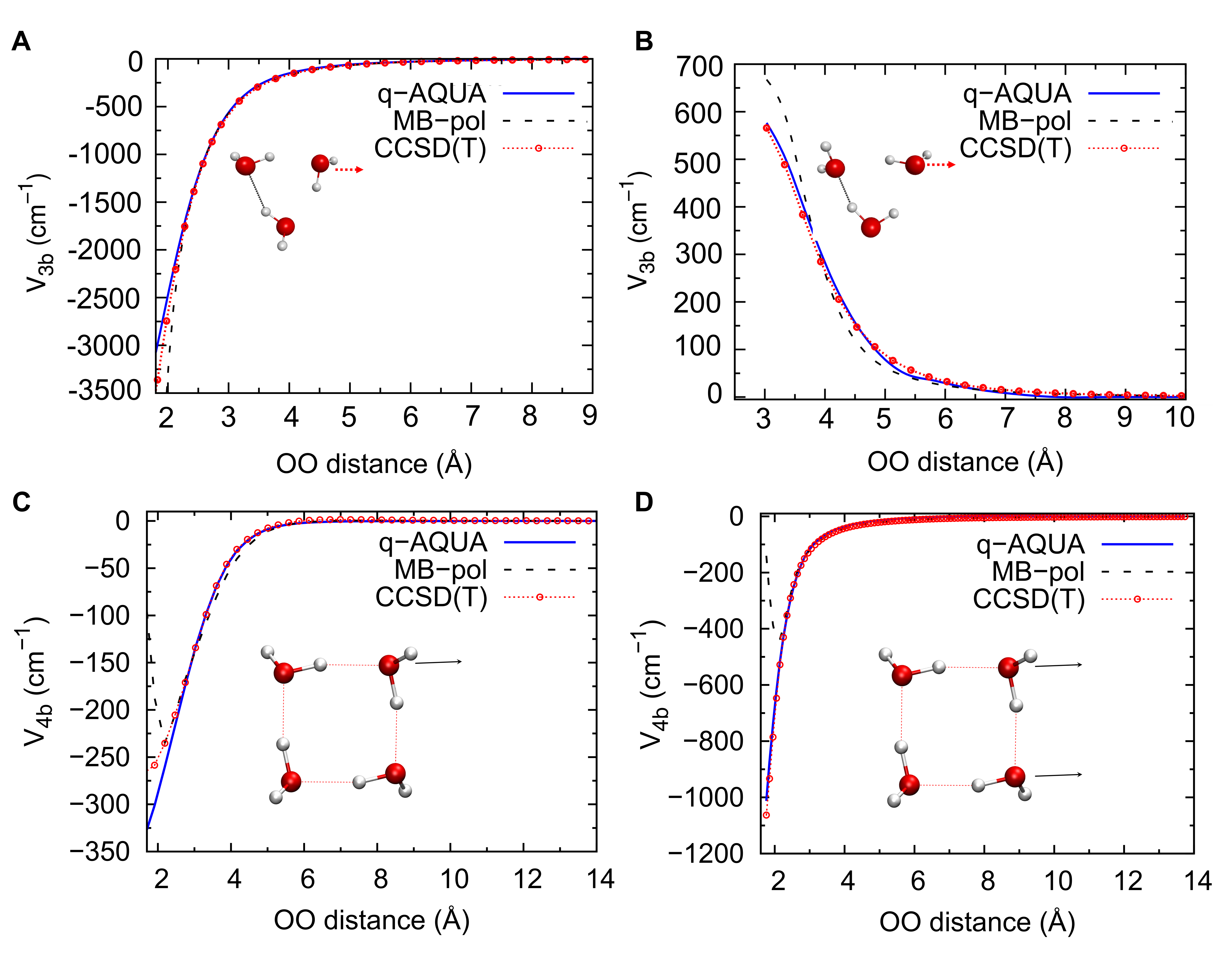}
\end{center}
\caption{Comparison of the new 3-b fit and MB-pol with direct CCSD(T) energies for an attractive cut (A) and an repulsive cut (B). Comparison of the new 4-b fit and direct CCSD(T) energies for a monomer-trimer cut (C) and a dimer-dimer cut (D).}
\label{fig:cuts}
\end{figure}

Panels A and B of Fig. \ref{fig:cuts} show attractive and repulsive cuts, respectively, for the 3-b potential as one monomer is moved relative to the remaining dimer. The q-AQUA potential provides excellent agreement with CCSD(T) calculations throughout the 2--10 \AA ~region. MB-pol is almost as accurate as q-AQUA for the attractive cut, but underestimates the repulsive 3-b potential in the 4--5.5 \AA ~range and overestimates it considerably when the OO distance is less than 4 \AA.   

Panel C of Fig. \ref{fig:cuts} shows a 4-b potential cut as one monomer is moved with respect to the remaining trimer.  The 4-b potential labeled ``MB-pol'' is the TTM4-F 4-b potential embedded in MB-pol. As seen, it is low  compared to the CCSD(T) calculations in the range of about 3--5 \AA, and is in strong disagreement with them below about 2.5 \AA.  The q-AQUA potential is in good agreement with CCSD(T) energies throughout the range shown. For moving one water dimer with respect to the other, panel D of Fig. \ref{fig:cuts} shows that, while the MB-pol has strong deviations from the CCSD(T) results below 2.5 \AA, the q-AQUA potential is in good agreement with the CCSD(T) results throughout the range. One might expect that the TTM4-F 4-body potential embedded in MB-pol would be uniformly accurate in the long range, but this is not the case.  This is shown in Fig. S9 of the SI, which plots the difference between CCSD(T)-F12 4-body energy and the MB-pol/TTM4-F 4-body energy against the maximum OO distance in the tetramer for all the configurations in our 4-body data set. The TTM4-F potential has large errors even when the OO distance is around 7 \AA. The RMSE for TTM4-F, as compared to the CCSD(T)-F12 benchmark is 21.2 cm$^{-1}$, whereas the RMSE for the q-AQUA 4-b potential against the same benchmark is 7.2 cm$^{-1}$. Note that the average absolute value of the 4-b energy in the data set is 31.9 cm$^{-1}$, so an RMSE of 21.2 cm$^{-1}$ from TTM4-F is large. A correlation plot between the q-AQUA 4-b energies and the CCSD(T)-F12 ones is provided in Fig. S10 of the SI, where additional precision metrics and properties of the present 2-, 3-, and 4-b PIP potentials can be found.

\begin{figure}[htbp!]
\begin{center}
\includegraphics[width=1.0\textwidth]{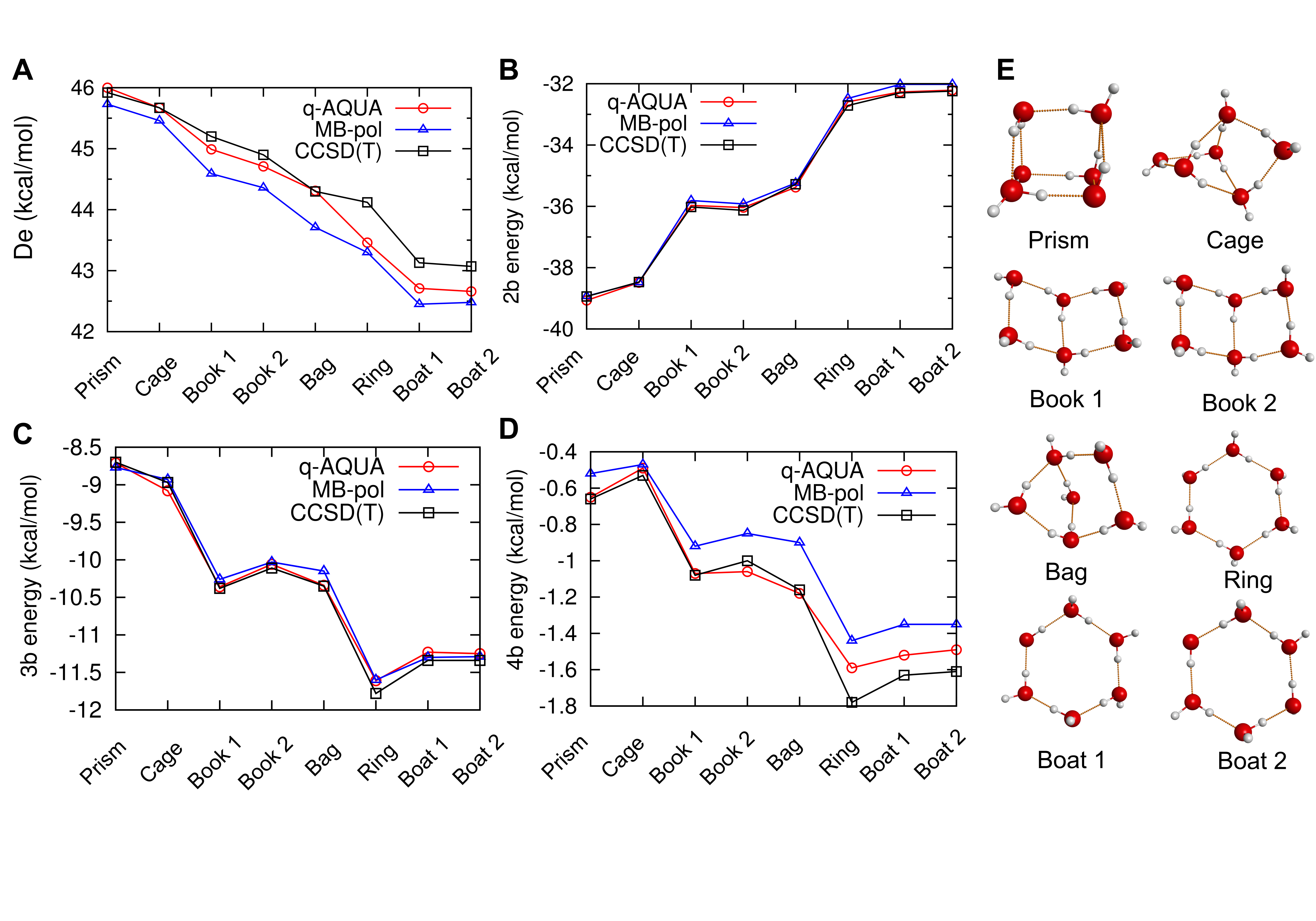}
\end{center}
\caption{Binding energies (A), 2-body energies (B), 3-body energies (C) and 4-body energies (D) for water hexamer isomers from present fits, MB-pol and benchmark CCSD(T) calculations (taken from refs. \citenum{bates09} and \citenum{mbpoltests} . (E) Structures of water hexamer isomers.}
\label{fig:hexamer}
\end{figure}

Next, we present tests of the accuracy of q-AQUA against benchmark results for the water hexamer and the 20-mer. Table S1 in the SI provides results for each of the 8 hexamer isomers, comparing the dissociation energies and the 2-b, 3-b, and 4-b energies for the q-AQUA potential, the MB-pol potential and the CCSD(T)/CBS calculations.\cite{bates09,mbpoltests}  The mean absolute errors (MAEs) are lower in general for the q-AQUA potential, although both appear to be fairly accurate. These results are shown graphically in Fig. \ref{fig:hexamer}, where different levels of agreement with the CCSD(T) results are seen. The D$_e$ results, particularly for the Ring, Boat1, and Boat2 isomers are more accurately predicted by q-AQUA than MB-pol, mostly because of differences in the 4-b contribution. Comparisons of the harmonic vibrational frequencies for four of the hexamer isomers are provided in Table S2 of the SI. As compared to the CCSD(T) benchmark calculation, q-AQUA and MB-pol are about equally good in predicting the frequencies of the Prism and Cage isomers, whereas q-AQUA does somewhat better than MB-pol on Book and Ring isomers.

The zero-point energies (ZPEs) of three hexamer isomers, Prism, Cage, and Book1, are calculated using the unrestricted diffusion Monte Carlo method.\cite{Anderson1975, Anderson1976, Schulten} Details of these calculations are provided in the SI. The ZPEs of the three isomers (all referenced to the electronic energy of the Prism equilibrium structure) using the full q-AQUA potential and without the 4-b contribution are listed in Table \ref{tab:dmc}, along with the statistical uncertainties. Note that due to a finite number of walkers and a finite step size, systematic errors on the absolute ZPE values exist, but early studies\cite{Mallory2015, wanghex} have shown that the energy differences between isomers are relatively insensitive to the number of walkers. The walker numbers used in this work are sufficient for a good estimate of the energy differences. As seen, the ZPE of the Cage is the lowest among the three isomers, by about 100 cm$^{-1}$ and thus the Cage is predicted to be the lowest energy isomer at 0 K.  This is in agreement with experiment\cite{clarycage,pate2012} and also the tentative conclusion of earlier DMC calculations using the WHBB PES.\cite{wanghex} and even earlier calculations using rigid-body DMC.\cite{clarycage} Further analysis of these DMC results and also those for larger clusters will be the subject of a future paper.

As noted, the DMC calculations are unconstrained, unlike studies using MB-pol, where geometric constraints are applied.\cite{Mallory2015, Mandprismhex} We have run unconstrained DMC calculations using MB-pol potential, and found many ``holes'', i.e., configurations with unphysical very negative energies. By contrast, q-AQUA is ``hole-free'' when running unconstrained DMC, and this finding was further corroborated by DMC runs on the 20-mer which we discuss briefly next. 

\begin{table}[htbp!]
\centering
\caption{Zero-point energies (in cm$^{-1}$) of three isomers of water hexamer from diffusion Monte Carlo calculations.}
\label{tab:dmc}

\begin{tabular*}{0.7\columnwidth}{@{\extracolsep{\fill}} l c c}
\hline
\hline
& With 4-b & Without 4-b \\
\hline
Prism & $32647 \pm ~9$ & $32598 \pm 12$ \\
Cage  & $32553 \pm 19$ & $32465 \pm ~9$ \\
Book 1  & $32652 \pm 12$ & $32740 \pm 16$ \\
\hline
\hline
\end{tabular*}
\end{table}

\begin{table}[htbp!]
\centering
\caption{Binding energies (in kcal/mol) of three \ce{(H2O)20} isomers of.}
\label{tab:dmc2}

\begin{threeparttable}
\begin{tabular*}{0.8\columnwidth}{@{\extracolsep{\fill}} l c c c c c}
\hline
\hline
Isomer & MP2/aV5Z & MP2/CBS & q-AQUA & MB-pol \\
\hline
A3 & -202.1 & $199.2 \pm{0.5}$\tnote{a} ($-200.8\pm{2.1}$\tnote{b}) & -199.8 &  -195.2 \\
A2d & -202.1 & n.a. & -201.7 & -195.3\\
9 & -201.5 & n.a. & -200.5 & -194.9 \\
\hline
\hline
\end{tabular*}

\begin{tablenotes}
\item[a] Ref. \citenum{heindeljpcl}
\item[b] CCSD(T)/CBS binding energy from Ref. \citenum{heindeljpcl}
\end{tablenotes}

\end{threeparttable}
\end{table}



Table \ref{tab:dmc2} shows the binding energies for three isomers of \ce{(H2O)20} and compares the benchmark MP2/aV5Z and MP2/CBS calculations\cite{heindeljpcl} with the predictions of q-AQUA and MB-pol. The q-AQUA prediction for the A3 isomer is in good agreement with the MP2/CBS value, whereas the MB-pol value is more than 4 kcal/mol too low.  Similarly, for the A2d and 9 isomers, the q-AQUA results are within one kcal/mol as compared to the available MP2/aV5Z results, while the MB-pol prediction is again to low by about 6--7 kcal/mol.  It is interesting to note that without the 4-b interaction the binding energies from q-AQUA are close to those from MB-pol and so the present 4-b interaction is needed to close the gap with the benchmark results. Preliminary DMC calculations for the 20-mer have been performed successfully and this validates the robustness of the q-AQUA potential for a large cluster.


Finally we examine the q-AQUA potential for simulations of bulk water properties. Specifically, classical molecular dynamics (MD), path integral molecular dynamics (PIMD), and ring polymer molecular dynamics (RPMD)\cite{qmdiffusion05,q-TIP4P,qmdiffusion05} were used to calculate both static and dynamic properties of liquid water. All the MD simulations were performed with the i-PI software,\cite{i-pi} and more computational details about calculations are provided in the SI. 

\begin{figure}[htbp!]
\begin{center}
\includegraphics[width=0.7\textwidth]{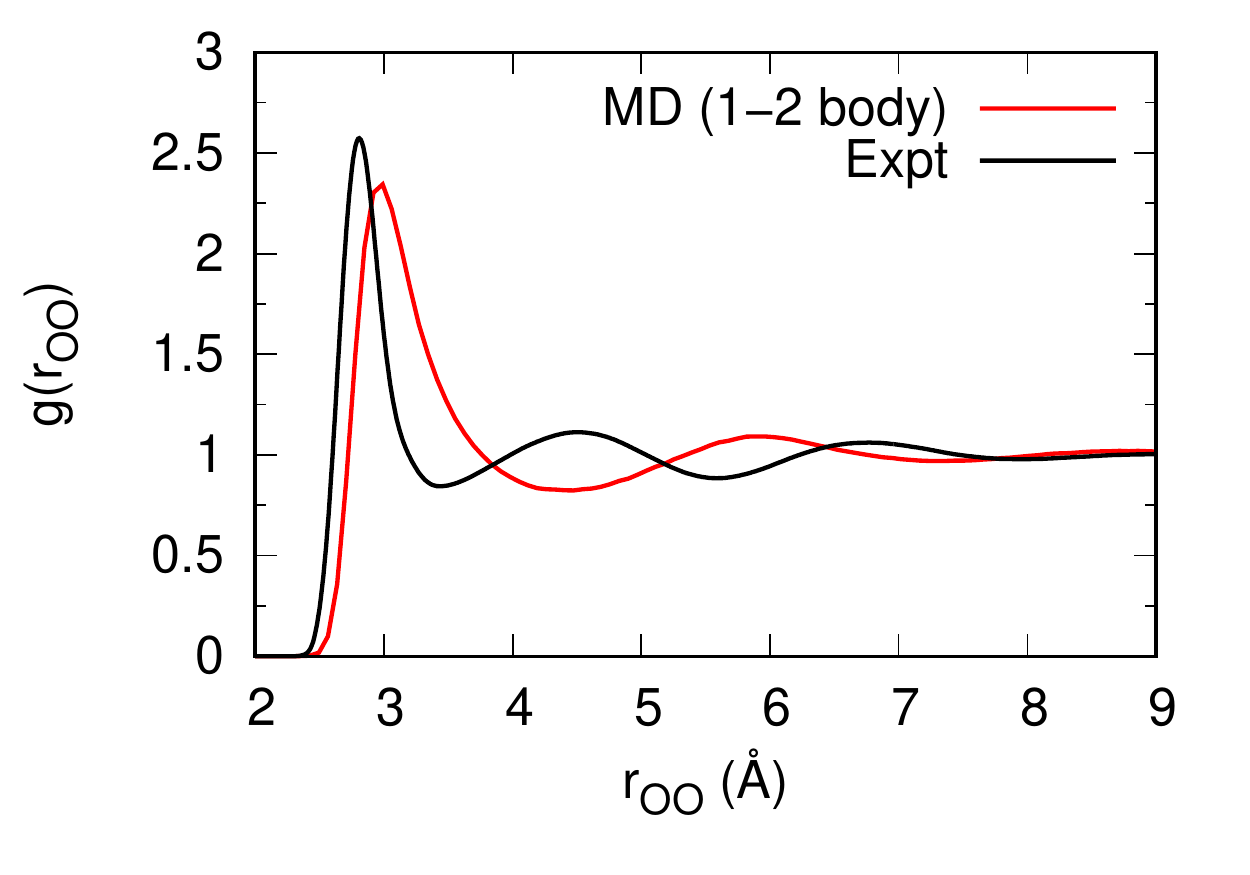}
\end{center}
\caption{OO radial distribution function from classical MD simulations at 298 K using reduced q-AQUA potential up to and including the 2-b interaction. The experimental data are from Ref. \citenum{Skinner2013}}
\label{fig:MD2b}
\end{figure}

\begin{figure}[htbp!]
\begin{center}
\includegraphics[width=1.0\textwidth]{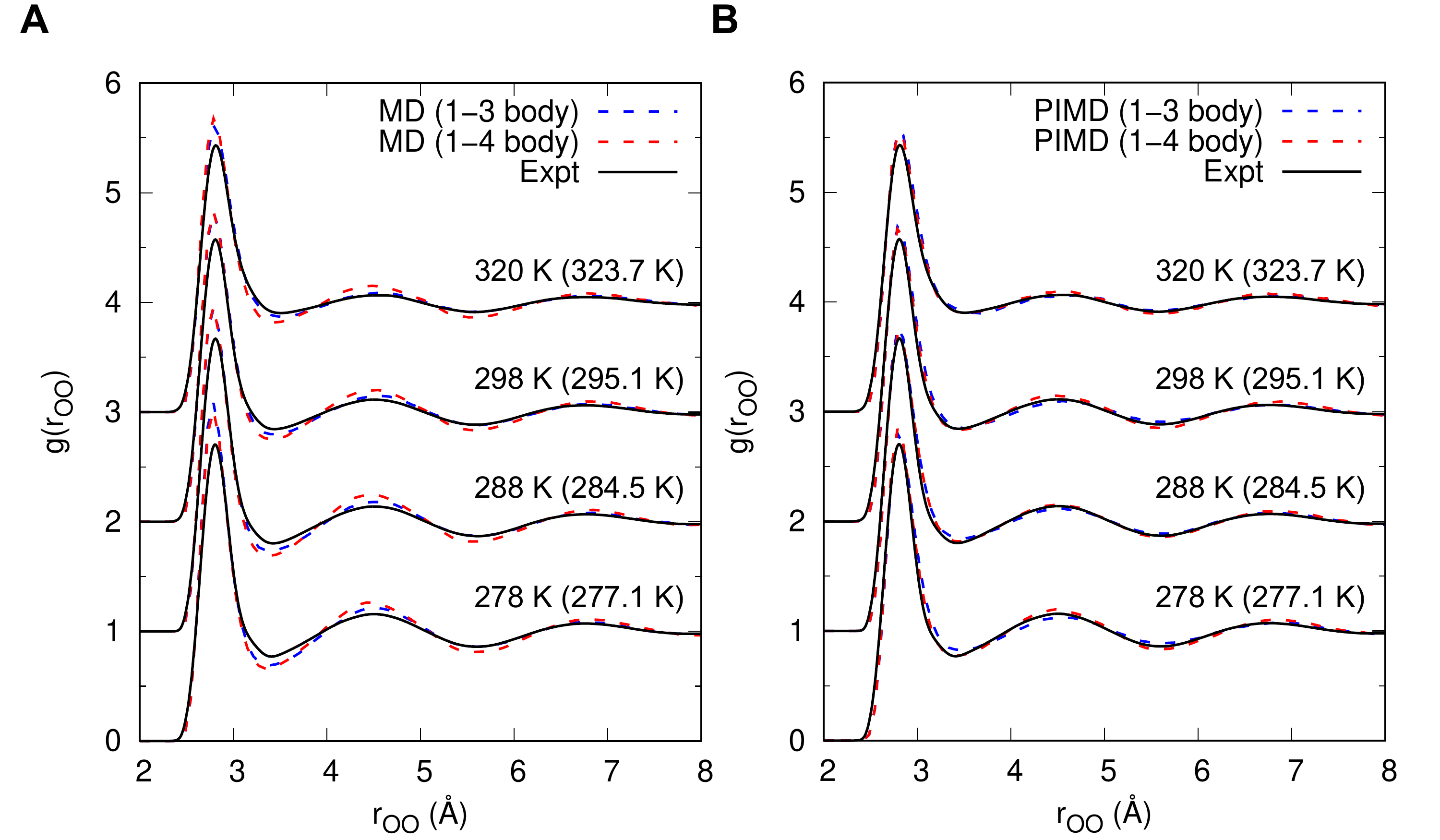}
\end{center}
\caption{OO radial distribution function from classical (A) and path integral (B) molecular dynamics simulations at different temperatures. The blue dashed lines are from reduced q-AQUA potential up to and including the 3-b interaction. The red dashed lines are from the full q-AQUA potential up to and including the 4-b interaction. The experimental data are taken from Ref. \citenum{Skinner2013,Skinner2014}}
\label{fig:MD}
\end{figure}

Fig. \ref{fig:MD2b} shows the OO radial distribution function obtained from classical molecular dynamics simulations at 298 K with only 1-b and 2-b interactions included. As seen, the simulated OO radial distribution deviates significantly from the experimental measurement. The inclusion of 1-b and 2-b interactions cannot sufficiently describe the water interactions in the condensed phase. Panels A and B of Fig. \ref{fig:MD} show the q-AQUA results for MD and PIMD simulations of the OO radial distribution function over a range of temperatures for 1-, 2-, 3-b and 1--4-body interactions compared to experiment (data taken from Refs. \citenum{Skinner2013, Skinner2014}).  Although the classical MD prediction agrees substantially with the position of the peaks in the distribution, it is noteworthy that the amplitudes of the peaks do not agree well. For the quantum calculations, the agreement is substantially better.  These plots do lead to the conclusion that for this property the 4-b interaction is not needed to obtain the graphical level of agreement seen. Fig. S14 in the SI shows this RDF using q-AQUA truncated at the 2-b, 3-b and 4-b levels.  As seen there, truncating at the 2-b level does not give an accurate result. Interestingly the peaks move to shorter OO distances in going from the 2-b, to 3-b and finally 4-b level of truncation.  This implies the presence of an effective additional attraction in going to the higher level of $n$-body interaction.

Next consider the self-diffusion constant as a function of temperature obtained with q-AQUA using MD and RPMD calculations. The results are given in Table \ref{tab:diffusion}. As seen, MD gets the trend correctly but is low compared to the experiment, whereas RPMD succeeds in coming close to the experiment. The increase in the self-diffusion constant obtained using RPMD is consistent with previous quantum mechanical calculations using the empirical q-TIP4/F water potential.\cite{q-TIP4P,qmdiffusion05}  The orientational relaxation time at 298 K is also given and, as seen, is in much better agreement when quantum RPMD as compared to the classical MD.  

Thus, for both the radial distribution function and especially the diffusion constant, quantum effects are significant for liquid water.  And it is clear that q-AQUA provides accurate results when coupled with quantum dynamics. 
Space does not permit a detailed discussion of these studies, but we simply note that the present calculations find significant quantum effects in both, and especially for the diffusion constant the magnitude seen here is consistent with recent results using an empirical water potential.\cite{RPMDiff2021} For the first time the effect of the 4-b has been shown to decrease the diffusion constant somewhat.  This implies that overall the 4-b is an added attraction which retards the diffusion, a result that seems quite reasonable to us. 

\begin{table}[htbp!]
\centering
\caption{Dynamical properties of liquid water from classical and quantum simulations with q-AQUA potential
\label{tab:diffusion}
}
\begin{threeparttable}
\begin{tabular*}{0.8\columnwidth}{@{\extracolsep{\fill}} c  c c c  }
\hline
\hline\noalign{\smallskip}
&\multicolumn{3}{c}{Self-diffusion coefficient,D,(\AA$^2$/ps)}   \\
Temperature (K) & Classical  & RPMD  &  Expt.$^a$   \\
\hline
278 & 0.080 $\pm$ 0.016 & 0.130 $\pm$ 0.015 & 0.131 \\
288 & 0.102 $\pm$ 0.017 & 0.177 $\pm$ 0.015 & 0.177 \\
298 & 0.145 $\pm$ 0.012 & 0.226 $\pm$ 0.020 & 0.230\\
320 & 0.248 $\pm$ 0.011 & 0.331 $\pm$ 0.016 & 0.360 \\
\hline\noalign{\smallskip}
&\multicolumn{3}{c}{Orientational relaxation time,$\tau_2$, (ps)}   \\
Temperature (K) & Classical  & RPMD  &  Expt.$^b$   \\
\hline
298 & 3.2 $\pm$ 0.1 & 2.4 $\pm$ 0.2 & 2.5 \\

\hline
\hline
\end{tabular*}
\begin{tablenotes}
\item[a] from Ref. \citenum{Mills1973} and \citenum{Holz2000}
\item[b] from Ref. \citenum{Bakker2005}
\end{tablenotes}
\end{threeparttable}
\end{table}


Finally, we present some timing results for q-AQUA. For the dynamics of 256 water molecules, Table S5 in the SI provides information for the time required for each of the $n$-body steps and for the total, with and without periodic boundary conditions (PBC), for one 2.4 GHz Intel Xeon core or eight using OpenMP, and for both the energy alone and for the energy and gradients.  The number of calculations required for each $n$-body component is also shown. The energy cost for the combined $n$-body components is about 2 s without PBC and about 4.5 s with PBC. The cost to get all gradients is about 2.3--2.4 times the cost of the energy due to the efficiency provided by the implementation of reverse derivatives.\cite{Houston2022} Using 8 cores rather than one speeds up the process by a factor of 6.2--6.6.  The most expensive part of the process is the calculation of the 4-body interactions, of which 268,304 or 115,922 are evaluated with or without PBC, respectively. For a single 200000-step MD trajectory with 256 water molecules under PBC, it takes around 40 hours when 15 CPU cores are used. For some calculations, it may be possible to truncate the MBE after the 3-b interactions, in which case the times are cut by more than a factor of two.

In summary, q-AQUA is a new water potential that is fully \textit{ab initio}-based and robust for quantum simulations.  An interesting aspect of this potential is that it can be used very efficiently at lower-levels of the many-body expansion and higher-body interactions can be investigated via approximate methods such as perturbation theory.  This opens up a new line of investigation that can be studied in the future.

\section{COMPUTATIONAL DETAILS}
The q-AQUA potential is composed of separate PIP fits for each of the $n$-body interactions with $n$=1--4.  The 1-b fit is the spectroscopically accurate water monomer PES calculated by Partridge and Schwenke.\cite{PS} The 2-b through 4-b fits are purified, compacted fits (2b is not purified) to new, expanded datasets containing CCSD(T) energies for 71892, 45332, and 3692 geometries, respectively.  The processes for purification and  compaction as well as for the addition of reverse derivatives to provide gradients are described in the SI, along with additional details for each of the fits.  

Briefly, the 2-b fit 
used a basis set of PIPs with 7th-order, 42 symmetry; it has an RMS error of 25 cm$^{-1}$.  The data set was limited to OO distances in the range from 2 to 8 \AA, whereas the long-range 2-b interaction was accounted for by a dipole-dipole interaction using a high-level dipole moment surface\cite{h2odip} and a smooth switching function.  

The new PIP 3-b PES that is a significant advance over the earlier one used in the WHBB PES. First, 45332 electronic energies are calculated at the CCSD(T)-F12a/aVTZ level of theory with BSSE correction included. This new 3-b data set extends over a broader energy range and covers a larger maximum OO distance range from 2.1 to 10.0 \AA. We then divided the new 3-b data set into two separate sets: one with maximum OO distance in the range from 2.0 to 7.0 \AA~with 42145 structures, and another with maximum OO distance in the range from 5.0 to 9.5 \AA, with 15282 structures. The short-range data set was fit using 4th-order 222111-symmetry PIPs of Morse variables, while the long-range data set is fit using 3rd-order 222111-symmetry PIPs of inverse internuclear distances. The fitting RMS errors for the short-range and long-range data sets are 9 cm$^{-1}$ and 11 cm$^{-1}$, respectively. A smooth switching function is used to join the two fits.

We recently reported the first CCSD(T)-based PES for the twelve-atom 4-b interaction,\cite{4b21} and the 4-b used in q-AQUA is an improved version of that PES. Here we just briefly describe the improvement. First, the size of the data set is expanded to 3692 from the original 2119 in order to cover more tetramer configurations. Second, we grouped the polynomials with the same coefficient into one polynomial, so that the data set does not have to be replicated 24 times for the fit, and the number of polynomials is greatly reduced. Lastly, the polynomial basis is augmented with a selection of 4th-order PIPs. The final basis set consists of 200 (grouped) polynomials and the RMS fitting error is 7.2 cm$^{-1}$. (For comparison, the RMS is 10.2 cm$^{-1}$ if the original basis is used to fit the expanded data set.)

Details of the hexamer results, the Diffusion Monte Carlo calculations and the MD and PIMD simulations are found in the SI.

\section{ASSOCIATED CONTENT}
\subsection*{Supporting Information available}

\begin{itemize}
\item Discussion of differences between the  q-AQUA, MB-pol, and WHBB potentials
\item Description of three methods to optimize the basis set of polynomials used in calculating$n$-body potential energy surfaces: Purification, Pruning, and Analytic Gradients  
\item Details concerning each of the 2-body, 3-body, and 4-body potential energy surfaces 
\item Further details concerning the hexamer results, including predictions of the q-AQUA and MB-pol potentials and comparison to the CCSD(T)/CBS calculations for the harmonic frequencies for four of the water hexamers
\item Details concerning the Diffusion Monte Carlo (DMC) calculations
\item Details concerning the Molecular Dynamics (MD) simulations 
\end{itemize}

\section{AUTHOR INFORMATION}
\subsection{Corresponding Authors}
*E-mail: q.yu@yale.edu\\
*E-mail: szquchen@gmail.com\\
*E-mail: PLH2@cornell.edu\\
*E-mail: riccardo.conte1@unimi.it\\
*E-mail: jmbowma@emory.edu\\
\subsection{ORCID}
Qi Yu: 0000-0002-2030-0671\\
Chen Qu: 0000-0001-8889-4851\\
Paul L. Houston: 0000-0003-2566-9539\\
Riccardo Conte: 0000-0003-3026-3875\\
Apurba Nandi: 0000-0002-6191-5584\\
Joel M. Bowman: 0000-0001-9692-2672\\
\subsection{Notes}
The authors declare no competing financial interest

\section*{ACKNOWLEDGEMENTS}
JMB thanks the ARO, DURIP grant (W911NF-14-1-0471), for funding a computer cluster where most of the calculations were performed and current financial support from NASA (80NSSC20K0360). QY thanks Professor Sharon Hammes-Schiffer and National Science Foundation (Grant No. CHE-1954348) for support.

\bibliography{refs}



\end{document}


\clearpage



\flushbottom
\maketitle

\thispagestyle{empty}
\clearpage
\section*{Summary}
This supplemental information first contains discussion of differences between the  q-AQUA, MB-pol, and WHBB potentials. Next we provide a description of three methods to optimize the basis set of polynomials used in calculating n-body potential energy surfaces: Purification, Pruning, and Analytic Gradients.  Details are then provided for each of the 2-body, 3-body, and 4-body potential energy surfaces.  The next section provides further details on the hexamer results, including comparison of the predictions of the q-AQUA and MB-pol potentials to the CCSD(T)/CBS calculations for the harmonic frequencies for four of the water hexamers. This is followed by sections providing details on the Diffusion Monte Carlo (DMC) calculations and on the Molecular Dynamics (MD) simulations.

\section*{Differences between q-AQUA, MB-pol, and WHBB potentials}

It is important to distinguish the new PES based on Eq. (1) of the main text from the MB-pol one, and the WHBB one.  First, we note that MB-pol is not strictly a many-body PES in the sense of Eq. (1) of the main text.  To see that, we give the explicit expression for the MB-pol potential for $N$ monomers.  It is 
\begin{equation}
  V_\text{MB-pol}(1,\cdots,N)=V_\text{TTM4-F}(1,\cdots,N)+\sum_{i>j}^N\Delta{V_{2b}(i,j)}S_{2b}+\sum_{i>j>k}^N\Delta{V_{3b}}(i,j,k)S_{3b},
\end{equation}
where $V_\text{TTM4-F}(1,\cdots,N)$ is the TTM4-F PES\cite{TTM4} and $\Delta V_{2b}(i,j)$ and $\Delta V_{3b}(i,j,k)$ are correction PESs. These are PIP fits to the difference of CCSD(T) and TTM4-F energies for 2-b and 3-b interactions, respectively. These corrections are relatively short range  and they are switched to zero at long range using switching functions, denoted generically by $S_{2b}$ and $S_{3b}$.  (Details of this switching are given in the MB-pol papers.\cite{mbpol2b,mbpol3b}) Note that  $V_\text{TTM4-F}(1,\cdots,N)$ contains the 1-b PS monomer potential\cite{PS} and also the PS dipole  moment surface as part of the electrostatic interaction. Higher-body interactions are given by polarization and electrostatic terms within the Thole formalism (some parameters in these used in MB-pol were optimized by Paesani and co-workers). $V_\text{TTM4-F}(1,\cdots,N)$ itself cannot be written simply as a many-body expansion.  The MB-pol approach is clearly very insightful; however, it does rely on the accuracy of the TTM4-F potential for 4-b and higher-body interactions for which no corrections are made.

By contrast the WHBB PES, which preceded the MB-pol one, is given by two variations. One is simply the MB expansion of Eq. (1) truncated at the 3-b level, namely
\begin{equation}
\begin{split}
V_\text{WHBB}(1,\cdots,N) = & \sum_{i=1}^NV_{1b}(i) + \sum_{i>j}^N[V^\text{CCSD(T)}_{2b}(i,j)S_{2b} + (1-S_{2b})V^\text{TTM3-F}_{2b}(i, j)] +\\
& \sum_{i>j>k}^NV_{3b}(i,j,k)S_{3b},
\end{split}
\end{equation}
where the monomer potential is the PS one, the 2-b term is a PIP fit to thousands of CCSD(T) energies, which is switched in the range 4.5--5.5 \AA  ~to the 2-b interaction from TTM3-F.\cite{WHBB,TTM3F} Details of this 2-b PES have been given previously.\cite{HBB2} Finally, a pure \textit{ab initio} 3-b interaction is a PIP fit to tens of thousands of MP2 energies.  This fit is switched to zero by an exponential damping function with a range of 8.0 bohr. Again details have been given previously.\cite{WHBB} An option in the WHBB PES is to add 4 and higher-body interactions using those in TTM3-F. However, this adds considerably to the cpu effort and makes only marginal improvement in benchmarks comparisons for the water hexamer.

\section*{Purification, Pruning, and Analytic Gradients}
\subsection*{The Purification Scheme}
Consider a group of identical particles whose potential energy surface we wish to describe using a multi-body expansion (MBE) (see Eq. (1) of the main text).  We wish to describe this potential energy surface by using a basis set of permutationally invariant polynomials (PIPs) whose coefficients will be determined by numerical regression so as to smoothly fit a dataset of electronic energies and, perhaps, gradients for different geometries and whose polynomials will be powers of, typically, Morse ($\exp(-r_{i,j}/a)$) or reciprocal variables of the internuclear distances.  Permutationally invariant polynomials have been discussed in detail previously.\cite{Braams2009,Bowman2010,ARPC2018}  It is convenient to generate them using the monomial symmetrization approach (MSA) software,\cite{Xie-nma,msachen} which ensures that the polynomials have the correct symmetry with respect to exchange of identical atoms.  However, there are two potential issues for application of MSA to the MBE problem. First, not all of these MSA PIPs will have the correct limiting behavior for the many body expansion. Second, the MSA software is not designed to consider the exchange of identical \textit{groups} of atoms, so we need to ensure this type of permutational symmetry by another method.

Consider first the issue of the limiting behavior. We use ``purification'' to refer to the process of identifying and setting aside PIPs that do not have the correct limiting behavior when individual monomers or groups of monomers are removed from the others to a great distance.  Because the fits are to the $n$-body contributions, when any of the group members are separated by large distances, thus eliminating the n-body nature of the interaction, the n-body contribution must go to zero. To ensure the correct limits for all polynomials, we take an arbitrary geometry, then remove $n$-mers from one another to large distances, and finally calculate the values of the PIPs to see whether they go to zero.  In the case of the 4-body interactions, we remove each of the 4 water monomers from the other three and each of the six pairs of water dimers from one another.  In the case of the 3-body interactions, we remove each water monomer from the remaining pair. 
In our calculations, the distances  are augmented by 100 \AA, and we accept the polynomial as having the correct behavior if its value is below $10^{-6}$.

Polynomials that do not have the correct limiting behavior cannot immediately be removed from consideration because there may be other polynomials that, for example, are composed of products of one with an incorrect limit and one with a correct limit.  Instead, the ones with an incorrect limit are simply given different names than those with the correct limit.  When the process of evaluating the polynomials is finished, we then look at the definitions of all those with the correct limits and determine which of the monomials and which of the renamed polynomials (with incorrect limits) are required to calculate them.  We then remove those that are not required and renumber those that remain, keeping the order of calculation so that no partial calculation that contributes to any polynomial needs to be performed twice.  The result is a set of polynomials that all have the correct limiting behavior and that can be calculated efficiently.

We now turn to the issue of permutational symmetry for exchange of the identical monomers.  As mentioned previously, when the monomers are groups of atoms, this exchange is not taken into account by the MSA software.  Thus, the polynomials that we create by purification will not, in general, have permutational symmetry with respect to exchange of identical monomers.  One method for dealing with this issue is to augment the dataset by adding all relevant permutations of the Cartesian coordinates and assigning them the same energy.  If there are $n$ monomers, this requires a set of $n!$ geometries for each energy.  A better method is to identify those groups of polynomials that have permutational symmetry with respect to monomer exchange and to then form ``superpolynomials'' that are the sum of the polynomial members of each group.  Each group will have $n!$, not necessarily unique, members whose sum is independent of the monomer permutations and which can then receive a single coefficient in the energy or gradient determination.  

We identify the permutationally invariant groups of polynomials by taking a single set of $n!$ permutationally related geometries and calculating the value of each polynomial.  Taken together, the groups of polynomials for each permutation will have the same group of values, but the values of individual polynomials will vary from permutation to permutation.  For each permutation, one can form pairs of the polynomial identities and their values, and then sort the pairs by their values.  Looking at all pairs that have the same value component in all permutations gives the identities of the polynomials, some of which may be repeated, that make up a permutationally invariant group. In general, there will be as many groups as there were original polynomials. These groups, each with $n!$ (not necessarily unique) polynomial contributions, are then summed to form the ``superpolynomials''.

\subsection*{Polynomial Pruning}
The MSA software referred to above is generated from just two parameters, the permutational symmetry of the atoms and the total order desired; i.e.,the maximum sum of the polynomial powers of the transformed internuclear distances.  The permutational symmetry need not be the complete symmetry of the system, but must include the relevant exchange possibilities.  For example, in the case of the water trimer, each water would have the symmetry designation of 21, meaning that the two hydrogens on the water can permute with one another and the oxygen does not permute.  The symmetry of the total system could be described as 63, but that would imply that the hydrogens or oxygen on one water could exchange with those on another.  At high enough energy, this rearrangement can certainly happen, but for lower energies, each water stays intact.  The advantage of not using 63 symmetry is that there are more relevant polynomials and coefficients, so the fit is generally more accurate for the energy range in which the waters remain intact.  Thus, one might consider, instead, the symmetry 222111, for which the hydrogens on any individual water can exchange with one another, but atoms on one water do not exchange with those any other water.  Entire water molecules can exchange, but only if this symmetry is dealt with by one of the methods discussed in the previous section.  

For any given permutational symmetry, the maximum number of PIPs will be determined by the maximum polynomial order desired.  Of course, the processes of purifying and grouping to achieve symmetric exchange of the waters as a whole will reduce the numbers of polynomials from that given by the MSA output.  In general and up to a limit, the more polynomials, the more accurate will be the fit, but the longer will be the time for calculating the energy and/or gradients. The limit is determined by the need to have far fewer coefficients than one has energy/gradient constraints so as to avoid ``overfitting'', a situation in which all the points are well-fit but the region between them has strong oscillations in energy. In many cases, one would like to have a number of polynomials/coefficients that is somewhere in between the numbers arrived at for two different choices of order.  In such cases, we ``prune'' the number of polynomials provided by the larger order to the number desired for a reasonable compromise between accuracy and speed.  We use the following method to decide which of the polynomials derived from the larger order should be kept.

A decision on which polynomials are most important may be made on the basis of the data set.  The first step is to evaluate the maximum values of the each transformed internuclear distance compared to the range of values among the geometries of the data set.  Recall that the internuclear distances are transformed, usually as Morse or reciprocal values.  Thus, long distances have small transformed values (and are less important) than short distances, which have transformed values nearer to unity (for the Morse transformation).  Taking the maximal values of the transformed internuclear distances, we then evaluate all the polynomials available from the MSA output for the larger order.  We then eliminate those polynomials with the smaller values until we arrive at the desired number of final polynomials.  The method works both for regular polynomials generated by the MSA software and for ``superpolynomials'' generated as described in the previous section. 

\subsection*{Analytical Gradients}
It is often very desirable to have analytical gradients of the potential so that forces between atoms can be calculated, either for use in the fitting process or, for example, in molecular dynamics simulations. Unfortunately, many higher levels of electronic structure theory do not automatically provide gradients without substantial execution cost.  There are several methods for accomplishing the goal of providing analytical gradients.  One is simply to differentiate all the terms in the energy expression with respect to each of the Cartesian coordinates.\cite{msachen}  However, this requires a time cost on the order of $3 N t_e$ for calculation of the gradients, where $N$ is the number of atoms and $t_e$ is the time for calculating the energy.  A better method is the reverse (or backward) derivative technique,\cite{BaydinPearlmutter2014,Baydin2018} which, for molecules up to 15 atoms, we have recently shown to have a time cost of 3--4 $t_e$, independent of the number of atoms.\cite{Houston2022}  We use this method for evaluating the gradients for all the n-body potentials described below.

\section*{Details concerning the Two-Body PES}
As pointed out in the literature numerous times and using numerous metrics, the 2-b interaction accounts for a substantial amount (around 80-90 \%) of the total interaction energy of water clusters.  Thus, it is important to obtain this interaction potential using the highest available level of electronic structure. CCSD(T) electronic energies for the 2-b interaction have been reported for flexible monomers  previously by us,\cite{HBB2} by Paesani and co-workers,\cite{mbpol2b} and more recently by Metz and Szalewicz.\cite{ksflexdimer} The data sets span significantly different energy ranges and configurations. For example our previous dataset of 30083 energies  spans an energy range predominantly from $\pm$ 7 kcal/mol with a significant number of configurations with energies as high as 28 kcal/mol. The MB-pol data set of about 40000 configurations is predominantly in the range $\pm$ 7 kcal/mol with a small number of configurations with energies as high as 14 kcal/mol. Finally the one from ref. \citenum{ksflexdimer} consists of 4758 energies and spans the range of -5 to 10 kcal/mol.


We generated a total of  71892 CCSD(T)/CBS 2-b energies with structures selected from our previous HBB2 data set as well as from the MB-pol data set, using the criteria that the  OO distances be in the range from 2.0 to 8.0 \AA. The 2-b energy is defined as:
\begin{equation}
    V_{2b} = V_{(\ce{H2O})_2} - V_{\ce{H2O},1}-V_{\ce{H2O},2}
\end{equation}
For each water dimer structure, we first performed CCSD(T)/aug-cc-pVTZ calculations with an additional 3s3p2d1f basis set, following the same procedure as MB-pol 2-b, namely, exponents equal to (0.9, 0.3, 0.1) for sp, (0.6, 0.2) for d, and 0.3 for f.  This additional basis is placed at the center of mass (COM) of each dimer configuration. We also determined the basis set superposition error (BSSE) correction with the counterpoise scheme. Second, we performed CCSD(T)/aug-cc-pVQZ calculations with the same additional basis set and with the BSSE correction. The final CCSD(T)/CBS energies were obtained by extrapolation over the CCSD(T)/aug-cc-pVTZ and CCSD(T)/aug-cc-pVQZ 2-b energies. All of these \emph{ab initio} calculations were performed using Molpro package.\cite{MOLPRO_brief} 

\begin{figure}[htbp!]
\begin{center}
\includegraphics[width=1.0\textwidth]{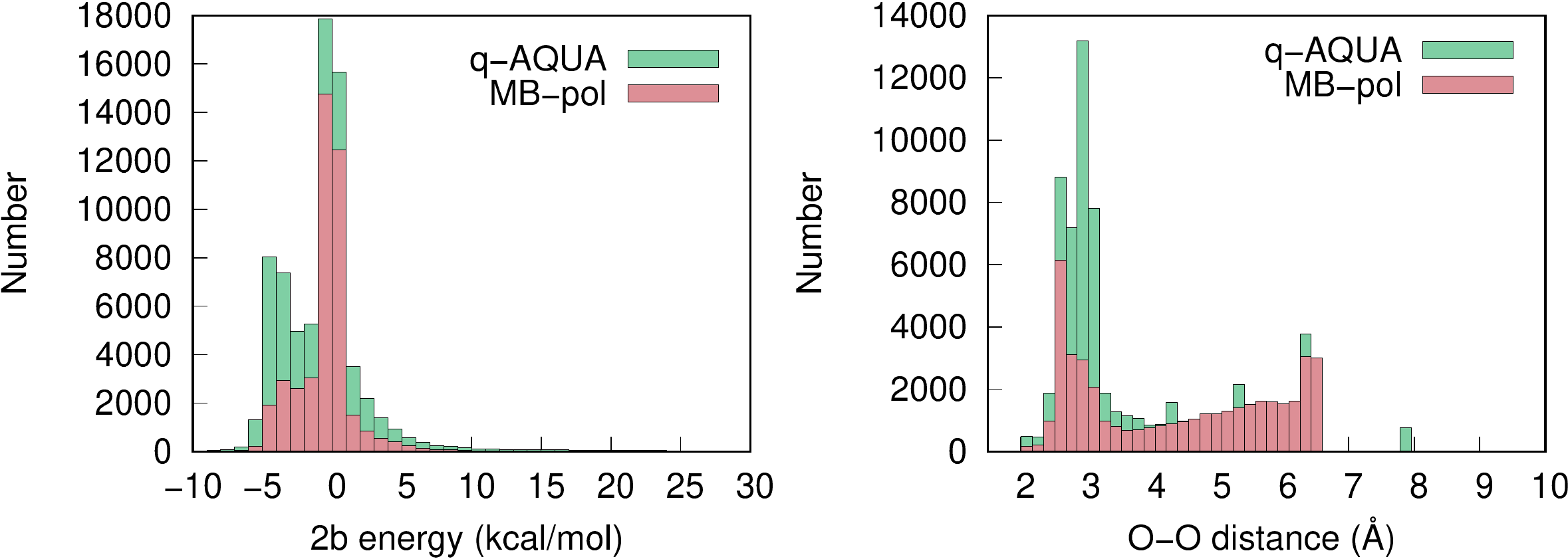}
\end{center}
\caption*{Figure S1: Distribution of 2-b energies.}
\end{figure}

Figure S1 shows the distribution of CCSD(T)/CBS 2-b energies in the range of -10 to 30 kcal/mol. Not shown are additional structures, numbering approximately 0.5\% of  the total, that have a high energy beyond this range. The OO distance distribution for the data set spans the range from 2 to 8 \AA. We fit a basis set of permutationally invariant polynomials to this new 2-b data set using 7th-order, 42 symmetry. The fitting RMS error for the whole data set is 25 cm$^{-1}$. Figure S2 shows the correlation plot between the 2-b fit and the CCSD(T)/CBS reference energies and confirms the accuracy of the final fit.  However, this 2-b data set only involves dimer structures with OO distance between 2 and 8 \AA,  and the long-range 2-body interaction is also important. To account for the long range 2-body interaction, we applied a dipole-dipole interaction potential using a high-level dipole moment surface.\cite{h2odip} This long-range 2-b interaction is expressed as:
\begin{equation}
    V_{2b}^{\text{long-range}} = \sum_{i=1}^3\sum_{j=1}^3\frac{q_iq_j}{r_{ij}}
\end{equation}
where $q_i,q_j$ are the partial charges on $i$th atom of monomer 1 and $j$th atom of monomer 2, and $r_{ij}$ is the distance between two atoms. The partial charge on each atom of the water monomer is obtained from the water monomer dipole moment surface by Lodi et al.\cite{h2odip}

\begin{figure}[htbp!]
\begin{center}
\includegraphics[width=0.8\textwidth]{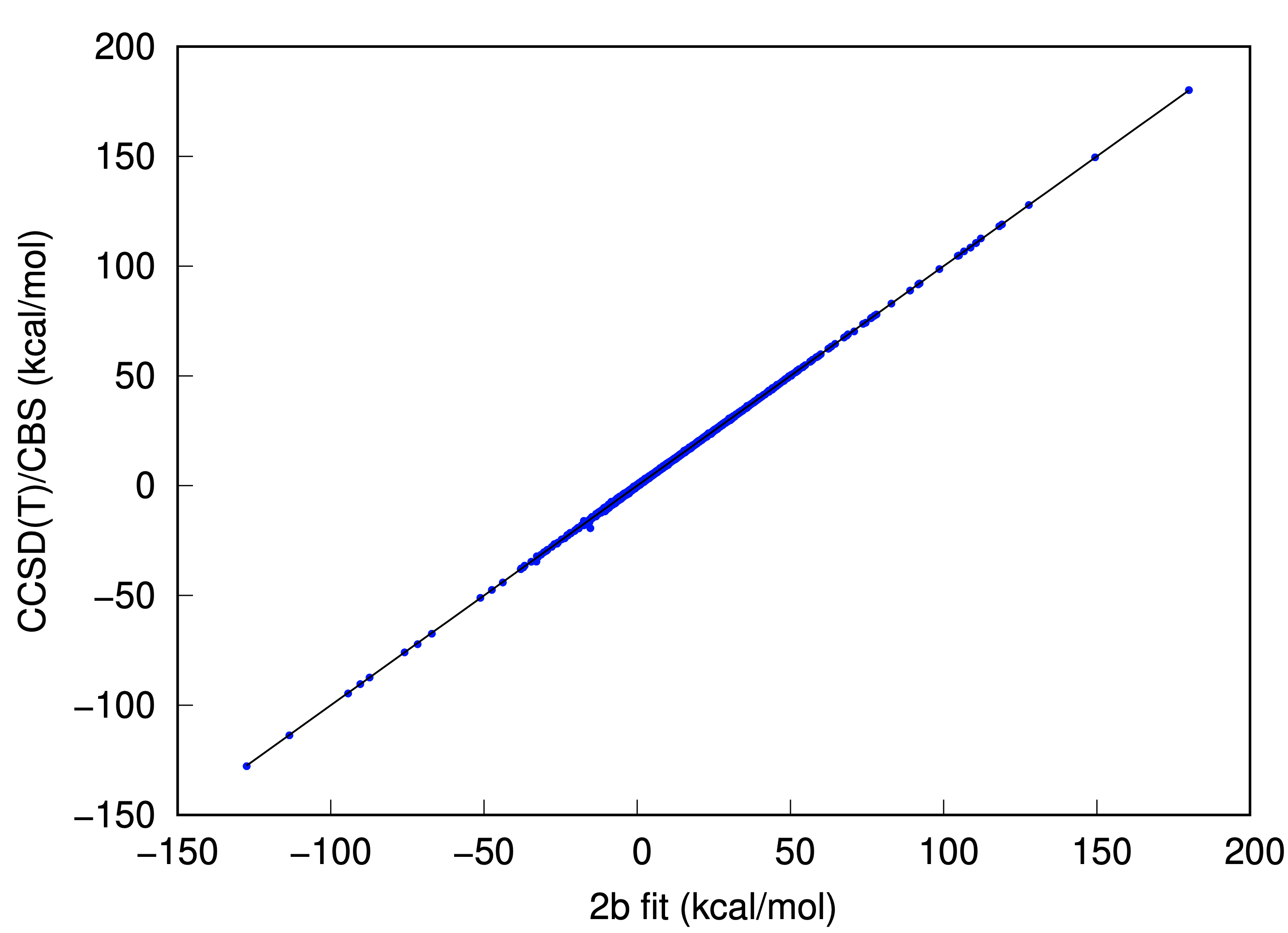}
\end{center}
\caption*{Figure S2: Correlation plot between 2-b fit and CCSD(T)/CBS reference data.}
\end{figure}

Finally, a smooth switching function was employed to connect the short-range 2-b fit to the long-range dipole-dipole interaction. The expression for the final 2-b potential energy surface is:
\begin{equation}
    V_{2b} = s(r_{\text{OO}})V_{2b}^{\text{fit}}+(1-s(r_{\text{OO}}))V_{2b}^{\text{long-range}}
\end{equation}
where $s$ is calculated as
\begin{equation*}
  \begin{aligned}
     s&=1.0,~~ r_{\text{OO}}\leq 6.5\ \text{\AA}\\
     &=1-10\left(\frac{r_{\text{OO}}-6.5}{7.8-6.5}\right)^3+15\left(\frac{r_{\text{OO}}-6.5}{7.8-6.5}\right)^4-6\left(\frac{r_{\text{OO}}-6.5}{7.8-6.5}\right)^5, ~~6.5<r_{\text{OO}}<7.8\ \text{\AA}\\
     &=0.0,  ~~r_{\text{OO}}\ge 7.8\ \text{\AA}
  \end{aligned}
\end{equation*}

Figures S3 and S4 show two potential cuts for the 2-b interaction and compare q-AQUA, MB-pol, and CCSD(T) results.

\begin{figure}[htbp!]
\begin{center}
\includegraphics[width=0.9\textwidth]{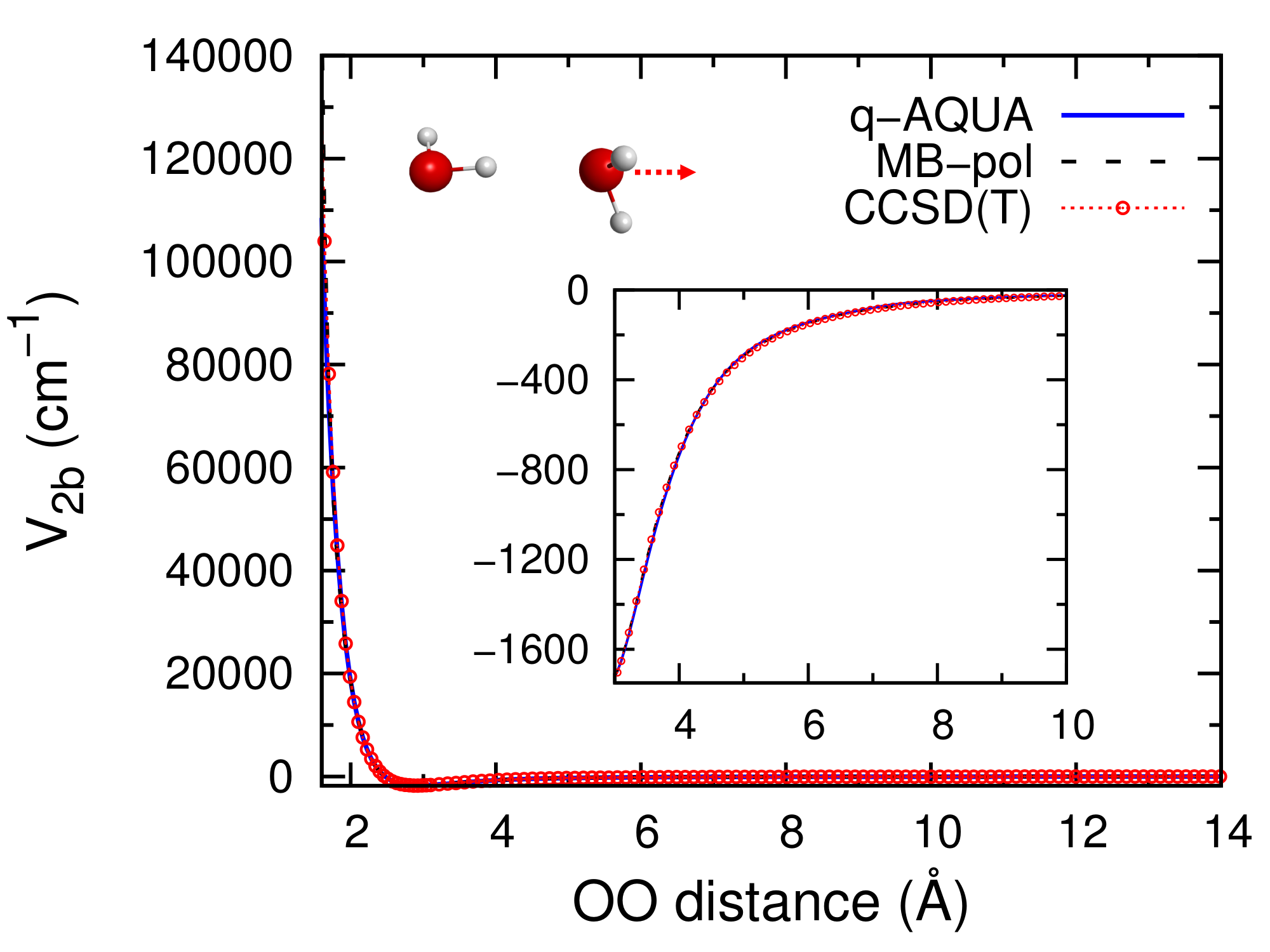}
\end{center}
\caption*{Figure S3: Comparison of the new 2-b fit and direct CCSD(T) energies for an attractive cut.}
\label{fig:2bcut1}
\end{figure}

\begin{figure}[htbp!]
\begin{center}
\includegraphics[width=0.9\textwidth]{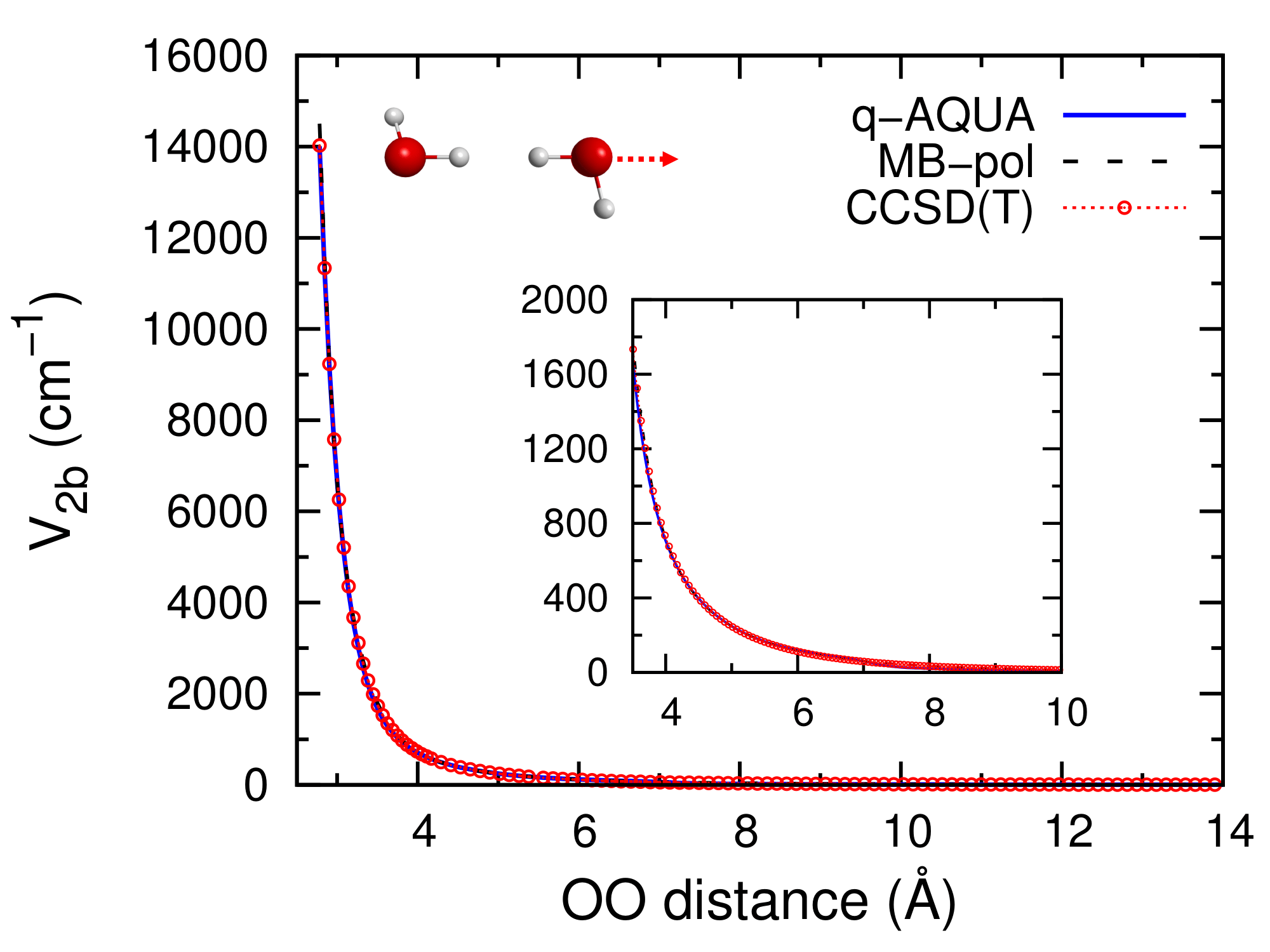}
\end{center}
\caption*{Figure S4: Comparison of the new 2-b fit and direct CCSD(T) energies for a repulsive cut.}
\label{fig:2bcut2}
\end{figure}

\section*{Details concerning the Three-Body PES}

 
We selected a total of 45332 trimer structures from the previous WHBB and MB-pol trimer configurations, using the criteria of keeping the maximum OO distance within the  2--9.5 \AA  ~range. The 3-b energy is defined as
\begin{equation}
    V_{3b} = V_{(\ce{H2O})_3} -\sum_{i}^3 V_{(\ce{H2O})_2,i}+\sum_{i}^3 V_{\ce{H2O},i}
\end{equation}
For each water trimer structure, CCSD(T)-F12a/aug-cc-pVTZ calculations were performed and the BSSE correction was included with the counterpoise scheme. All calculations were conducted using the Molpro package.\cite{MOLPRO_brief} The distributions of the 3-b energies and the maximum OO distances are shown in Figure S5. For the calculated 3-b energies, most trimer structures have 3-b interaction within the range of -6 to 6 kcal/mol, as can be seen in the left panel of Fig. S5. For the maximum OO distance distribution, it can be seen from the right panel in Figure S5 that current data set spans a significantly broader region than the MB-pol data set. 

We divided the new 3-b data set into two separate sets: one with maximum OO distance in the range from 2.0 to 7.0 \AA, and another with maximum OO distance in the range from 5.0 to 9.5 \AA, resulting in two data sets with 42145 and 15282 structures, respectively. The short-range data set was fit using 4th-order 222111-symmetry permutational invariant polynomials which are functions of Morse variables, $\exp(-r_{ij}/a)$, where $r_{ij}$ is the internuclear distance and $a=3$ bohr. The fitting RMS error for this data set is 9 cm$^{-1}$. The long-range data set is fit using 3rd-order 222111-symmetry permutational invariant polynomials, which are functions of inverse of the internuclear distance, $r_{ij}$. The fitting RMS error for this data set is 11 cm$^{-1}$. The correlation plots between two 3-b fits and BSSE corrected CCSD(T)-F12a/aug-cc-pVTZ reference data are shown in Figs. S6 and S7, respectively. 

\begin{figure}[htbp!]
\begin{center}
\includegraphics[width=1.0\textwidth]{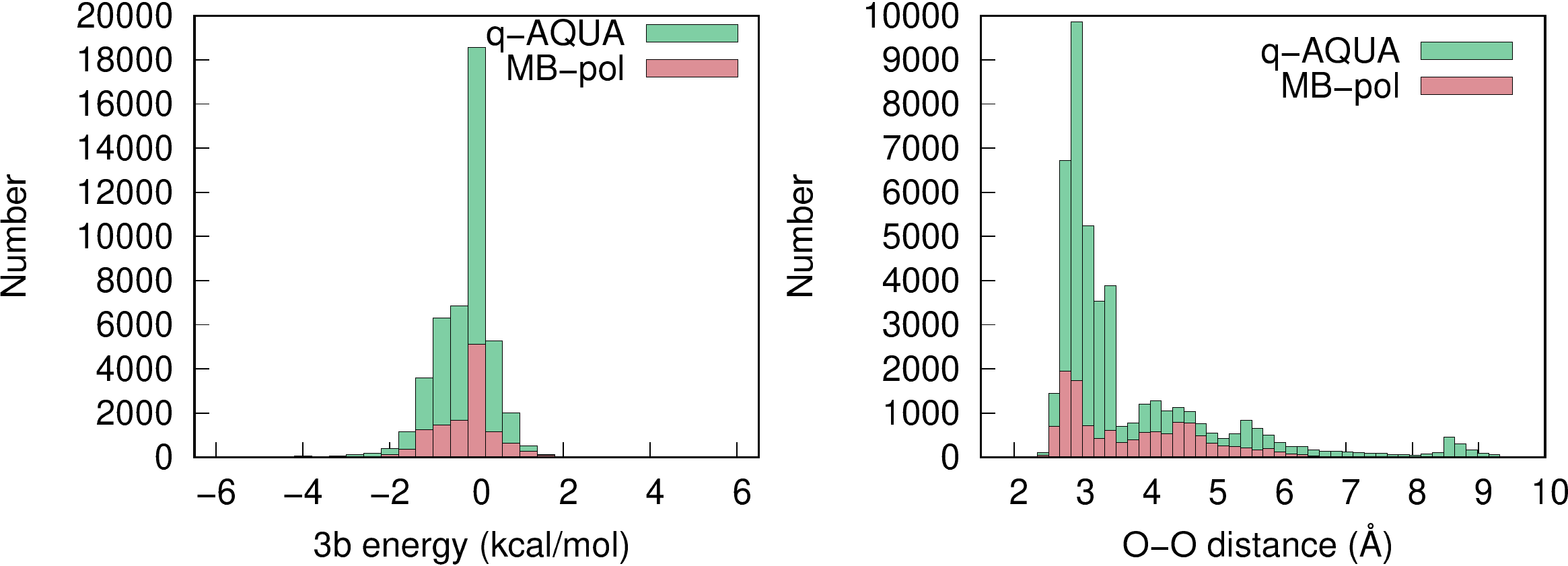}
\end{center}
\caption*{Figure S5: Distribution of 3-b energies, left panel, and OO distances for MB-pol and q-AQUA data sets.}
\end{figure}

Our 3-b potential energy surface is a mixed one with the two separate fits using a switching function based on the maximum OO distance, and also smoothly switched to zero. The final expression is :
\begin{equation}
    V_{3b} = s_1(r_{\text{OO}})\times\left[s_2(r_{\text{OO}})V_{3b}^{\text{short-range}}+(1-s_2(r_{\text{OO}}))V_{3b}^{\text{long-range}}\right]
\end{equation}
where $s_1$ is calculated by
\begin{equation*}
  \begin{aligned}
     s_1&=1.0,~~ r_{\text{OO}}\leq 7.0\ \text{\AA}\\
     &=1-10\left(\frac{r_{\text{OO}}-7.0}{9.0-7.0}\right)^3+15\left(\frac{r_{\text{OO}}-7.0}{9.0-7.0}\right)^4-6\left(\frac{r_{\text{OO}}-7.0}{9.0-7.0}\right)^5, ~~7.0<r_{\text{OO}}<9.0\ \text{\AA}\\
     &=0.0, ~~ r_{\text{OO}}\ge 9.0\ \text{\AA}
  \end{aligned}
\end{equation*}
and $s_2$ is calculated by
\begin{equation*}
  \begin{aligned}
     s_2&=1.0, ~~r_{\text{OO}}\leq 5.0\ \text{\AA}\\
     &=1-10\left(\frac{r_{\text{OO}}-5.0}{6.0-5.0}\right)^3+15\left(\frac{r_{\text{OO}}-5.0}{6.0-5.0}\right)^4-6\left(\frac{r_{\text{OO}}-5.0}{6.0-5.0}\right)^5,~~ 5.0<r_{\text{OO}}<6.0\ \text{\AA}\\
     &=0.0, ~~ r_{\text{OO}} \ge 6.0\ \text{\AA}
  \end{aligned}
\end{equation*}

\begin{figure}[htbp!]
\begin{center}
\includegraphics[width=0.8\textwidth]{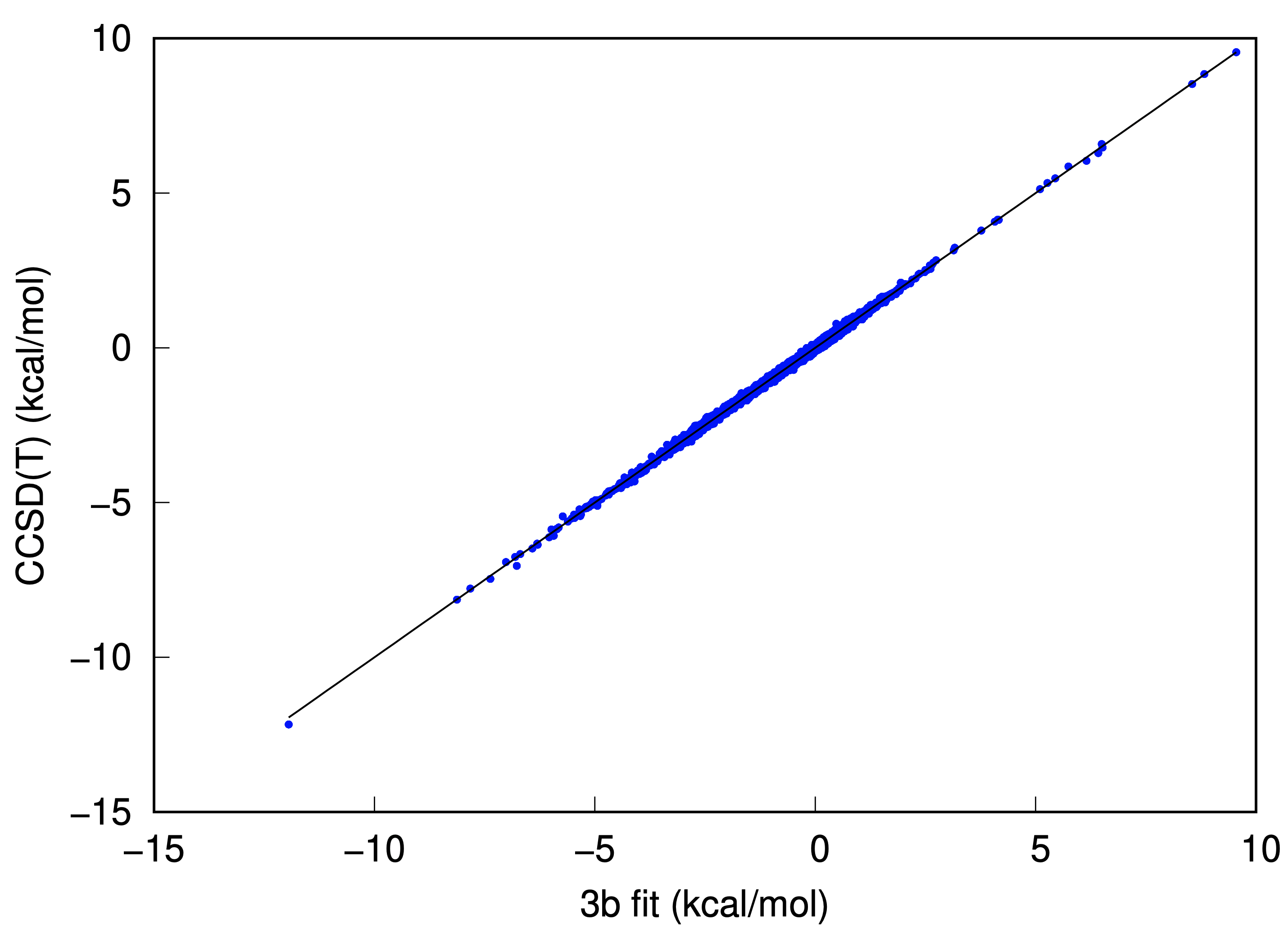}
\end{center}
\caption*{Figure S6: Correlation plot between short-range 3-b fit and BSSE-corrected CCSD(T)-F12a/aVTZ reference data.}
\end{figure}

\begin{figure}[htbp!]
\begin{center}
\includegraphics[width=0.8\textwidth]{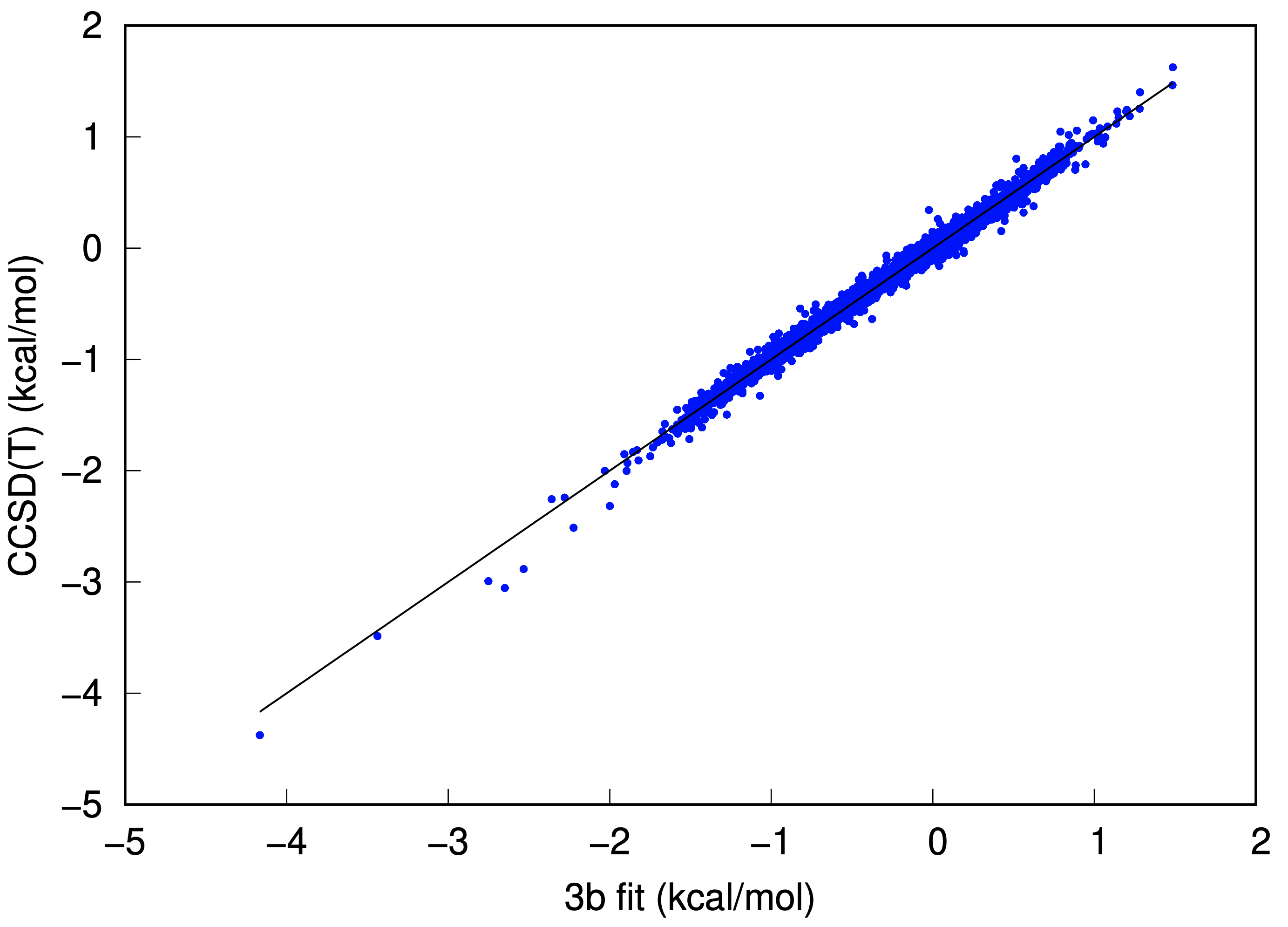}
\end{center}
\caption*{Figure S7: Correlation plot between long-range 3-b fit and BSSE-corrected CCSD(T)/CBS reference data.}
\end{figure}

\begin{figure}[htbp!]
\begin{center}
\includegraphics[width=0.8\textwidth]{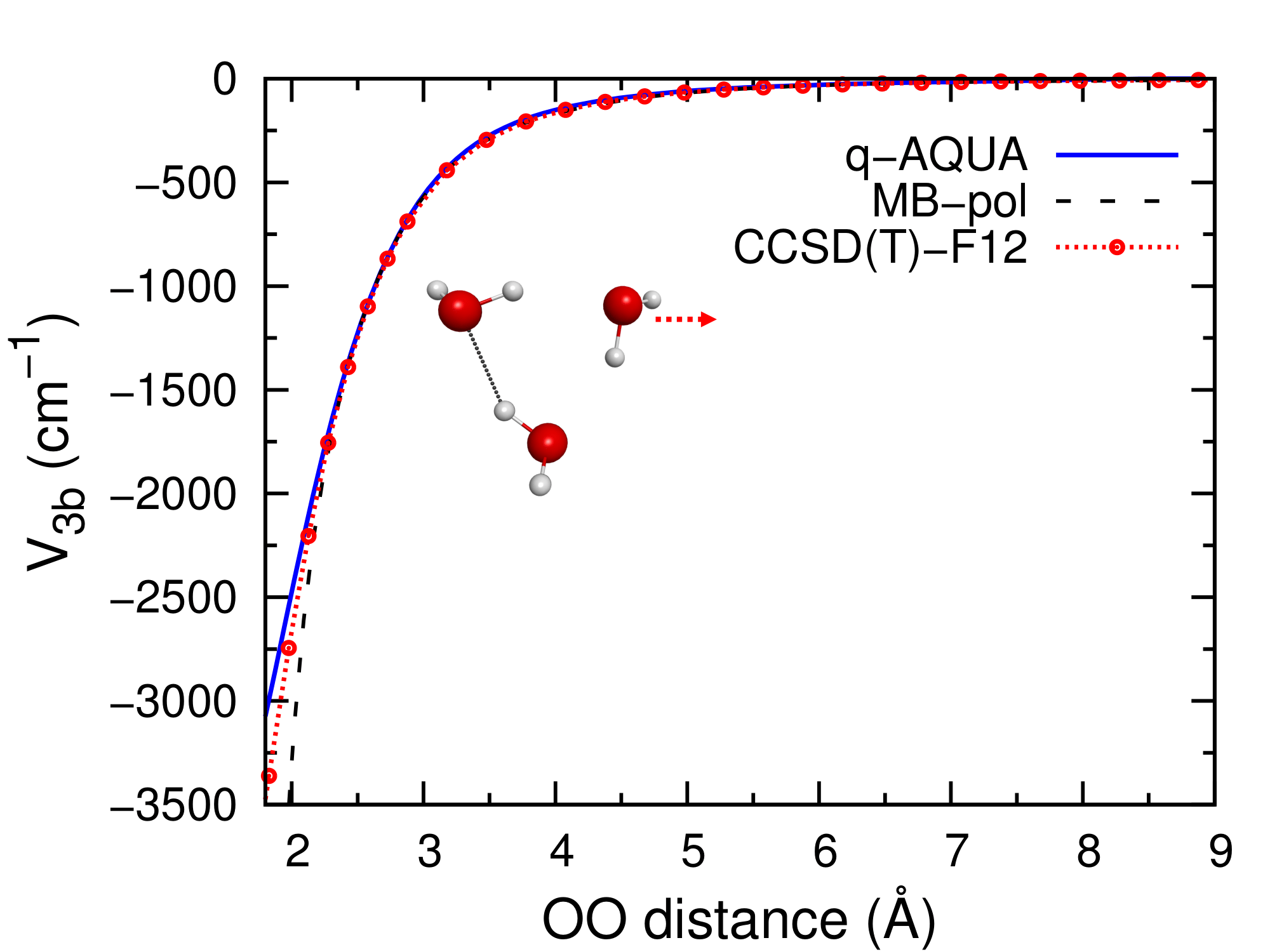}
\end{center}
\caption*{Figure S8: Comparison of the new 3-b fit and direct CCSD(T) energies for an attractive cut.}
\label{fig:3bcut1}
\end{figure}

Figure S8 shows an attractive cut through the potential as one monomer is moved relative to the remaining dimer.  The results of the q-AQUA and MB-pol potentials are shown along with the calculated CCSD(T)-F12 results.  Both potentials do well in the range down to about 2.5 \AA, while the MB-pol potential slightly underestimates and the q-AQUA sightly overestimates the potential at shorter distances.  Results for a repulsive cut have been shown in the main text.

\clearpage

\section*{Details concerning the Four-Body PES}

The four-body potential energy surface was a new potential which improved on the one previously reported.\cite{4b21}  It was based on 4th-order 222221111 symmetry.  The basis set was purified to ensure the correct limiting behavior at long distances between each water monomer and the remaining trimer and between each pair of water dimers.  Compaction reduced the number of monomials, polynomials, and ``renamed'' polynomials to 930, 1648, and 279, respectively, from the original basis set of 2910 monomials and 10,736 polynomials.  Determining the groups of polynomials that were permutationally symmetric with respect to water exchange further reduced the number of polynomials to 872 while increasing the number of renamed polynomials to 1927.  The advantages of the grouping are that an unreplicated data set can be used in the fit and that the fitting time is reduced. In order to achieve a good balance between speed and accuracy, the number of polynomials was further reduced to 200 by the pruning method described above.  The Fortran program produced for evaluating energies also used the reverse derivative method for calculating gradients.  

The coefficients were determined by fitting the basis set to a set of 3692 geometric points calculated at the CCSD(T)-F12 level.  Figure S9 shows the distributions of the 4-b energy and the OO distances from the data base.  Figure S10 provides a correlation plot between the CCSD(T) energies and the fit energies. The RMS fitting error is 7.2 cm$^{-1}$.

\begin{figure}[htbp!]
\begin{center}
\includegraphics[width=1.0\textwidth]{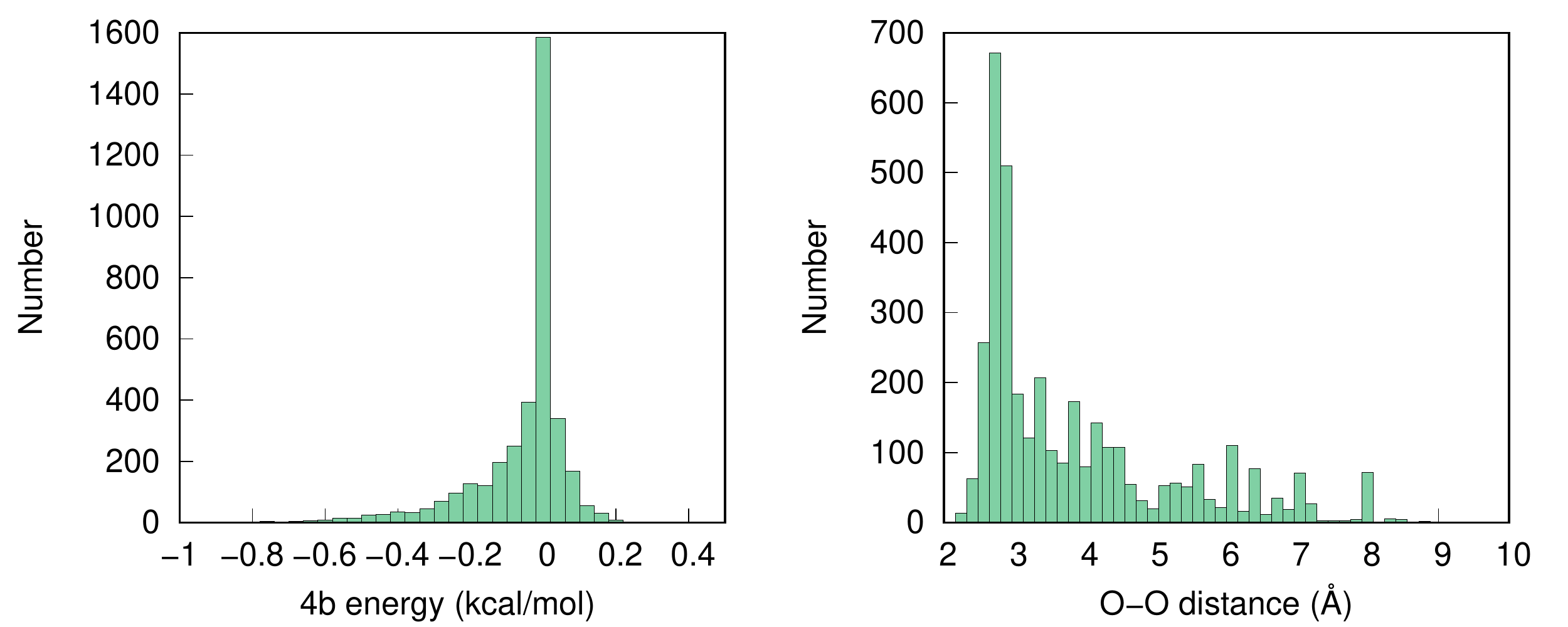}
\end{center}
\caption*{Figure S9: Distribution of 4-b energies and OO distances.}
\end{figure}

\begin{figure}[htbp!]
\begin{center}
\includegraphics[width=0.9\textwidth]{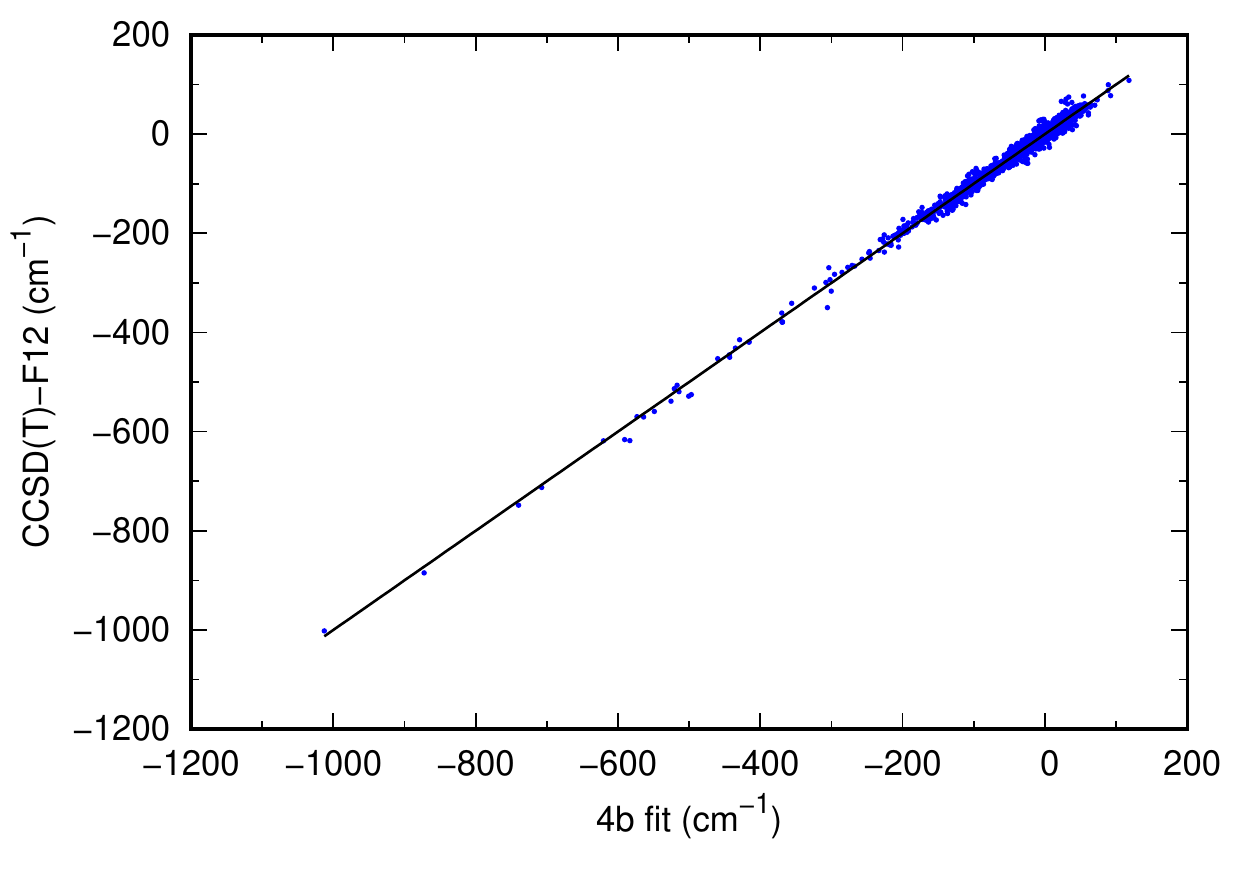}
\end{center}
\caption*{Figure S10: Correlation plot between 4-b fit and CCSD(T)-F12/haTZ references.}
\end{figure}

One might expect that MB-pol 4-body, which uses the TTM4-F, should be precise in the long range, but this is not the case, as is shown in Fig. S11. In this figure, the difference between CCSD(T)-F12 4-body energy and the MB-pol/TTM4-F 4-body energy is plotted against the maximum OO distance in the tetramer, for all the 3692 configurations in our 4-body data set. We can see that TTM4-F has large errors even when the OO distance is around 7 \AA. The RMSE of the TTM4-F 4-body is 21.2 cm$^{-1}$; for comparison, our 4-body fit has an RMSE of 7.2 cm$^{-1}$.

\begin{figure}[htbp!]
\begin{center}
\includegraphics[width=0.9\textwidth]{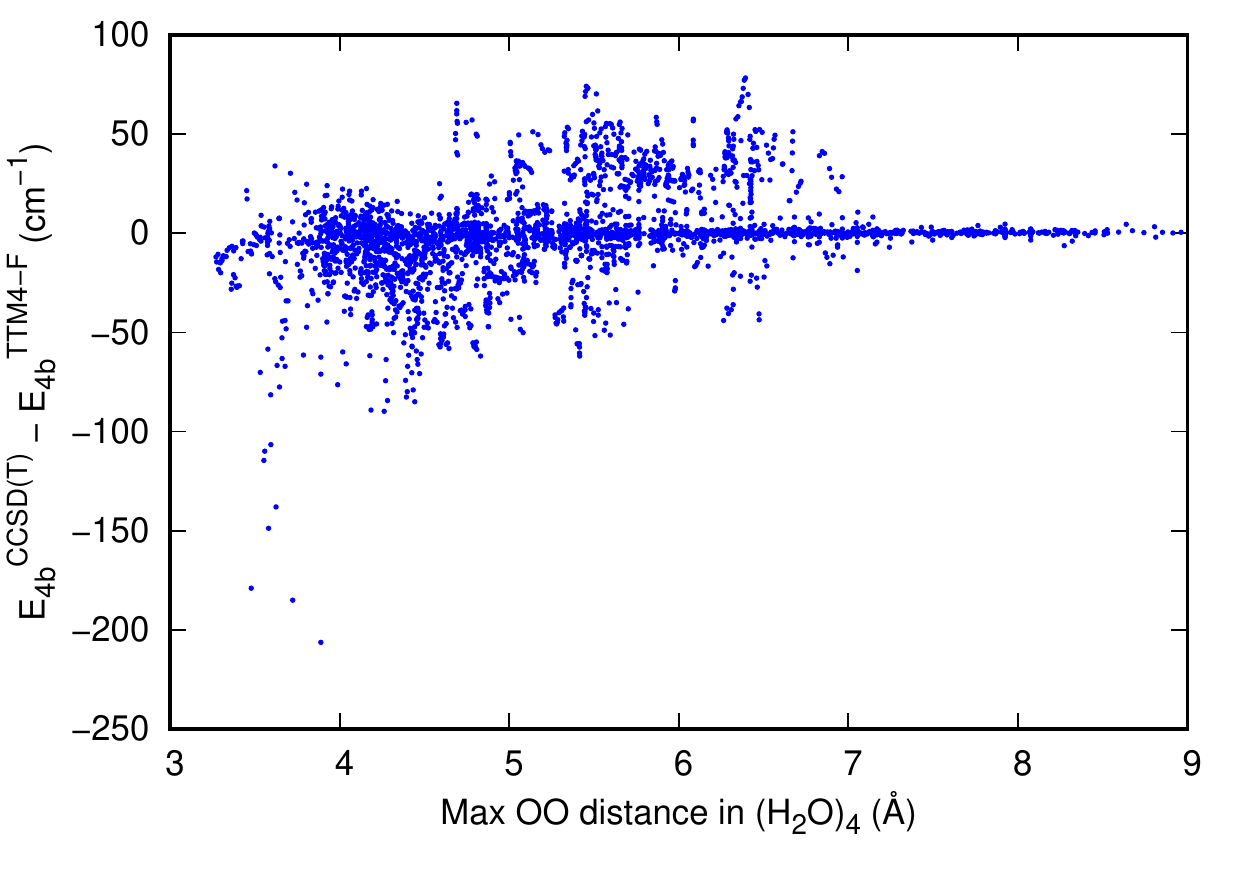}
\end{center}
\caption*{Figure S11: Difference between CCSD(T) 4-body energy and TTM4-F 4-body energy, as a function of the maximum OO distance in the tetramer.}
\end{figure}

\clearpage
\section{Details concerning water hexamers}
\subsection{Table of 2-b, 3-b, 4-b, and total dissociation energies for water hexamer isomers}
Table S1 shows results for the water hexamers (a plot of these data is given in the main text).

\begin{table}[htbp!]
\centering
\caption*{Table S1: 2-b, 3-b, 4-b and total dissociation energies (kcal/mol) for various water hexamer isomers.
\label{tab:tab_nmers}
}
\begin{threeparttable}
\begin{tabular*}{\columnwidth}{@{\extracolsep{\fill}} l | c c c | c c c }

\hline
\hline
 & \multicolumn{3}{c|}{De}  & \multicolumn{3}{c}{2b energy}   \\
\hline
Isomer & CCSD(T)/CBS\tnote{a}  & q-AQUA  &  MB-pol &CCSD(T)/CBS\tnote{b}  & q-AQUA  &  MB-pol  \\
\hline
Prism &   45.92 & 46.00 &  45.73 & -38.94 & -39.07 & -38.93 \\
Cage &   45.67 & 45.67 &  45.46 & -38.47 & -38.49 & -38.48\\
Book 1 &   45.20 & 44.99 &  44.59 & -36.02 & -35.97 & -35.81\\
Book 2 &   44.90 & 44.71 &  44.36  & -36.13 & -36.04 & -35.92\\
Bag &   44.30 & 44.31 &  43.71 & -35.28 & -35.38 & -35.24 \\
Ring &   44.12 & 43.46 &  43.30  & -32.71 & -32.59 & -32.48\\
Boat 1 &   43.13 & 42.71 &  42.45  & -32.30 & -32.27 & -32.02\\
Boat 2 &   43.07 & 42.66 &  42.48  & -32.24 & -32.21 & -32.02\\
\hline
MAE &  /  & 0.25 &  0.53 & / & 0.07 & 0.15\\
\hline
 & \multicolumn{3}{c|}{3-b energy}  & \multicolumn{3}{c}{4-b energy}   \\
\hline
Isomer & CCSD(T)/CBS\tnote{b}  & q-AQUA  &  MB-pol &CCSD(T)/CBS\tnote{b}  & q-AQUA  &  MB-pol  \\
\hline
Prism &   -8.70 & -8.71 &  -8.77 & -0.66 & -0.65 & -0.52 \\
Cage &   -8.97 & -9.08 &  -8.93 & -0.53 & -0.49 & -0.47\\
Book 1 &   -10.38 & -10.36 &  -10.26 & -1.08 & -1.07 & -0.92\\
Book 2 &  -10.11 & -10.06 &  -10.03  & -1.00 & -1.06 & -0.85\\
Bag &   -10.35 & -10.34 &  -10.15 & -1.16 & -1.18 & -0.90 \\
Ring &   -11.78 & -11.61 &  -11.60  & -1.78 & -1.59 & -1.44\\
Boat 1 &  -11.34 & -11.23 &  -11.30  & -1.63 & -1.52 & -1.35\\
Boat 2 &  -11.34 & -11.25 &  -11.29  & -1.61 & -1.49 & -1.35\\
\hline
MAE &  /  & 0.07 &  0.10 & / & 0.07 & 0.21\\
\hline
\hline
\end{tabular*}

\begin{tablenotes}
\item[a] From Ref. \citenum{bates09}
\item[b] Ref. \citenum{mbpoltests}
\end{tablenotes}

\end{threeparttable}
\end{table}

\subsection{Harmonic Frequencies of Water Hexamers}
Tables S2 and S3 show a comparison of the harmonic frequencies for four water hexamers (Prism, Cage, Book1, and Ring) for the q-AQUA and MP-pol potentials as compared with the benchmark CCSD(T)/CBS calculations of Ref. \citenum{Howard2015}.  As can be seen, the results for q-AQUA and MB-pol are both fairly accurate, with a slight improvement in the mean absolute error (MAE) from the q-AQUA potential.  For both q-AQUA and MB-pol, the discrepancies between the potential energy fits and the CCSD(T)/CBS calculations are dominated by the high frequency modes.

\begin{table}[htbp!]
\footnotesize
\caption*{Table S2: Comparison of harmonic frequencies (in cm$^{-1}$) of Prism and Cage hexamers from q-AQUA and MB-pol with CCSD(T)/CBS frequencies.\textsuperscript{a}}
\label{tab:freq1}

\begin{threeparttable}
	\begin{tabular*}{\columnwidth}{@{\extracolsep{\fill}}| l | c | c | c | c | c | c |}
	\hline
	Isomer & \multicolumn{3}{c |}{Prism} & \multicolumn{3}{c |}{Cage} \\
    \hline
    Method & CCSD(T) & q-AQUA & MB-pol & CCSD(T) & q-AQUA & MB-pol \\
    \hline
    mode 1 &   61 &   61 &   62 &   42 &   42 &   43 \\
    mode 2 &   70 &   71 &   73 &   56 &   58 &   57 \\
    mode 3 &   74 &   78 &   75 &   74 &   75 &   75 \\
    mode 4 &   98 &   91 &  110 &  100 &   99 &   98 \\
    mode 5 &  112 &  103 &  120 &  127 &  120 &  129 \\
    mode 6 &  149 &  145 &  150 &  153 &  149 &  153 \\
    mode 7 &  173 &  168 &  173 &  185 &  173 &  178 \\
    mode 8 &  178 &  175 &  179 &  194 &  186 &  191 \\
    mode 9 &  212 &  211 &  204 &  210 &  209 &  207 \\
    mode 10 &  217 &  225 &  218 &  223 &  218 &  214 \\
    mode 11 &  238 &  234 &  230 &  231 &  227 &  223 \\
    mode 12 &  246 &  241 &  241 &  234 &  234 &  230 \\
    mode 13 &  275 &  279 &  272 &  242 &  244 &  238 \\
    mode 14 &  284 &  282 &  277 &  253 &  249 &  247 \\
    mode 15 &  287 &  287 &  281 &  281 &  277 &  267 \\
    mode 16 &  357 &  346 &  347 &  293 &  295 &  290 \\
    mode 17 &  367 &  362 &  366 &  383 &  386 &  378 \\
    mode 18 &  420 &  417 &  407 &  395 &  399 &  391 \\
    mode 19 &  427 &  422 &  424 &  437 &  437 &  426 \\
    mode 20 &  462 &  462 &  447 &  453 &  456 &  441 \\
    mode 21 &  491 &  488 &  490 &  467 &  463 &  461 \\
    mode 22 &  530 &  530 &  527 &  534 &  528 &  523 \\
    mode 23 &  547 &  541 &  536 &  553 &  539 &  542 \\
    mode 24 &  612 &  604 &  590 &  620 &  618 &  603 \\
    mode 25 &  638 &  636 &  625 &  682 &  683 &  671 \\
    mode 26 &  675 &  669 &  661 &  717 &  720 &  700 \\
    mode 27 &  711 &  709 &  696 &  774 &  761 &  752 \\
    mode 28 &  823 &  821 &  808 &  790 &  785 &  772 \\
    mode 29 &  868 &  859 &  851 &  852 &  842 &  833 \\
    mode 30 & 1001 &  998 &  977 &  975 &  973 &  952 \\
    mode 31 & 1663 & 1658 & 1652 & 1666 & 1660 & 1663 \\
    mode 32 & 1674 & 1667 & 1671 & 1673 & 1667 & 1672 \\
    mode 33 & 1683 & 1678 & 1680 & 1684 & 1681 & 1681 \\
    mode 34 & 1699 & 1696 & 1689 & 1698 & 1694 & 1691 \\
    mode 35 & 1716 & 1709 & 1711 & 1707 & 1704 & 1704 \\
    mode 36 & 1733 & 1730 & 1729 & 1722 & 1716 & 1712 \\
    mode 37 & 3301 & 3317 & 3306 & 3324 & 3339 & 3310 \\
    mode 38 & 3509 & 3519 & 3512 & 3517 & 3530 & 3530 \\
    mode 39 & 3601 & 3620 & 3603 & 3556 & 3556 & 3563 \\
    mode 40 & 3620 & 3651 & 3627 & 3604 & 3632 & 3616 \\
    mode 41 & 3717 & 3753 & 3730 & 3650 & 3675 & 3671 \\
    mode 42 & 3735 & 3762 & 3748 & 3718 & 3759 & 3736 \\
    mode 43 & 3784 & 3812 & 3792 & 3757 & 3784 & 3768 \\
    mode 44 & 3799 & 3824 & 3807 & 3792 & 3819 & 3806 \\
    mode 45 & 3821 & 3835 & 3819 & 3895 & 3894 & 3901 \\
    mode 46 & 3898 & 3898 & 3909 & 3896 & 3905 & 3905 \\
    mode 47 & 3899 & 3908 & 3911 & 3899 & 3909 & 3912 \\
    mode 48 & 3901 & 3913 & 3912 & 3908 & 3928 & 3916 \\
    \hline
    MAE &  /  &  7.9 &  7.8 &   /  &  7.7 &  8.9 \\   
	\hline
   
	\end{tabular*}
   
   \begin{tablenotes}
   \item[a] From Ref. \citenum{Howard2015}
   \end{tablenotes}
  
\end{threeparttable}
\end{table}

\begin{table}[htbp!]
\footnotesize
\caption*{Table S3: Comparison of harmonic frequencies (in cm$^{-1}$) of book-1 and ring hexamers from q-AQUA and MB-pol with CCSD(T)/CBS frequencies.\textsuperscript{a}}
\label{tab:freq2}

\begin{threeparttable}
	\begin{tabular*}{\columnwidth}{@{\extracolsep{\fill}}| l | c | c | c | c | c | c |}
	\hline
	Isomer & \multicolumn{3}{c |}{Book-1} & \multicolumn{3}{c |}{Ring} \\
    \hline
    Method & CCSD(T) & q-AQUA & MB-pol & CCSD(T) & q-AQUA & MB-pol \\
    \hline
    mode  1 &   27 &   24 &   25 &   28 &   24 &   27 \\
    mode  2 &   37 &   37 &   36 &   28 &   24 &   27 \\
    mode  3 &   53 &   52 &   51 &   45 &   41 &   43 \\
    mode  4 &   67 &   67 &   69 &   45 &   41 &   43 \\
    mode  5 &   85 &   86 &   86 &   50 &   45 &   48 \\
    mode  6 &  156 &  155 &  155 &   82 &   77 &   77 \\
    mode  7 &  179 &  176 &  172 &  156 &  155 &  152 \\
    mode  8 &  189 &  183 &  181 &  172 &  161 &  167 \\
    mode  9 &  195 &  194 &  189 &  195 &  184 &  182 \\
    mode 10 &  225 &  219 &  214 &  195 &  184 &  182 \\
    mode 11 &  233 &  229 &  223 &  211 &  205 &  200 \\
    mode 12 &  245 &  238 &  232 &  211 &  205 &  200 \\
    mode 13 &  250 &  247 &  236 &  254 &  241 &  236 \\
    mode 14 &  271 &  269 &  257 &  254 &  241 &  236 \\
    mode 15 &  282 &  275 &  266 &  282 &  268 &  263 \\
    mode 16 &  291 &  287 &  277 &  292 &  286 &  276 \\
    mode 17 &  302 &  301 &  289 &  292 &  286 &  276 \\
    mode 18 &  377 &  369 &  367 &  323 &  315 &  305 \\
    mode 19 &  393 &  383 &  381 &  407 &  387 &  386 \\
    mode 20 &  432 &  422 &  426 &  426 &  404 &  409 \\
    mode 21 &  443 &  437 &  437 &  426 &  404 &  409 \\
    mode 22 &  467 &  458 &  458 &  441 &  432 &  431 \\
    mode 23 &  533 &  525 &  522 &  450 &  434 &  438 \\
    mode 24 &  601 &  592 &  586 &  450 &  434 &  438 \\
    mode 25 &  708 &  698 &  678 &  757 &  737 &  728 \\
    mode 26 &  735 &  731 &  711 &  776 &  753 &  738 \\
    mode 27 &  811 &  798 &  777 &  776 &  753 &  738 \\
    mode 28 &  829 &  820 &  800 &  867 &  842 &  818 \\
    mode 29 &  874 &  866 &  835 &  867 &  842 &  818 \\
    mode 30 &  989 &  980 &  953 &  941 &  916 &  892 \\
    mode 31 & 1661 & 1654 & 1655 & 1665 & 1665 & 1659 \\
    mode 32 & 1673 & 1669 & 1667 & 1676 & 1670 & 1668 \\
    mode 33 & 1675 & 1672 & 1670 & 1676 & 1670 & 1668 \\
    mode 34 & 1691 & 1682 & 1683 & 1701 & 1691 & 1691 \\
    mode 35 & 1702 & 1696 & 1695 & 1701 & 1691 & 1691 \\
    mode 36 & 1730 & 1721 & 1721 & 1716 & 1706 & 1705 \\
    mode 37 & 3386 & 3397 & 3413 & 3440 & 3469 & 3488 \\
    mode 38 & 3455 & 3474 & 3463 & 3505 & 3534 & 3531 \\
    mode 39 & 3503 & 3540 & 3521 & 3505 & 3534 & 3531 \\
    mode 40 & 3587 & 3598 & 3612 & 3554 & 3595 & 3577 \\
    mode 41 & 3637 & 3661 & 3653 & 3554 & 3595 & 3577 \\
    mode 42 & 3651 & 3680 & 3667 & 3570 & 3622 & 3596 \\
    mode 43 & 3768 & 3791 & 3781 & 3901 & 3901 & 3908 \\
    mode 44 & 3893 & 3896 & 3899 & 3901 & 3901 & 3908 \\
    mode 45 & 3898 & 3905 & 3908 & 3901 & 3901 & 3908 \\
    mode 46 & 3900 & 3907 & 3909 & 3901 & 3902 & 3911 \\
    mode 47 & 3900 & 3909 & 3912 & 3901 & 3902 & 3911 \\
    mode 48 & 3903 & 3916 & 3913 & 3901 & 3907 & 3912 \\
    \hline
    MAE &  /  &  8.2 & 12.6 &   /  & 13.5 & 16.5 \\   
	\hline
   
	\end{tabular*}
   
   \begin{tablenotes}
   \item[a] From Ref. \citenum{Howard2015}
   \end{tablenotes}
  
\end{threeparttable}
\end{table}

\section{Details of diffusion Monte Carlo calculations}
The use of the diffusion Monte Carlo (DMC) method to estimate the vibrational ground-state energy (i.e., zero-point energy) is based on the similarity between the diffusion equation and the imaginary-time Schr\"odinger equation with an energy shift $E_\text{ref}$
\begin{equation}
    \frac{\partial \psi(\bm{x},\tau)}{\partial \tau} = 
    \sum_{i=1}^{N}\frac{\hbar^2}{2m_i}\nabla_i^2\psi(\bm{x},\tau)
    -\left[ V(\bm{x})-E_{\text{ref}}\right]\psi(\bm{x},\tau)
\end{equation}
The reference energy $E_\text{ref}$ in the above equation is used to stabilize the diffusion system in its ground state and thus is the estimator of the zero-point energy.\cite{Anderson1975} 
We employed the unbiased, unconstrained implementation of DMC, \cite{Schulten} in which the DMC calculation starts from an initial guess of the ground-state wave function, represented by a population of $N(0)$ equally weighted Gaussian random walkers. These walkers then diffuse randomly in imaginary time according to a Gaussian distribution and the population is controlled by birth-death processes, given as
\begin{align}
    & P_\text{birth}= \exp[-(E_i-E_\text{ref}) \Delta\tau] - 1{\ \ \ } (\text{if}{\ } E_i < E_\text{ref}),\\
    & P_\text{birth}= 1 - \exp[-(E_i-E_\text{ref}) \Delta\tau]{\ \ \ } (\text{if}{\ } E_i > E_\text{ref}),
\end{align}
where $E_i$ is the energy of the $i$th walker. To maintain the number of random walkers at about the initial value $N(0)$, $E_\text{ref}$ is adjusted at the end of each time step according to
\begin{equation}
    E_\text{ref}(\tau) = \langle V(\tau) \rangle - \alpha\frac{N(\tau)-N(0)}{N(0)}
\end{equation}
where $N(\tau)$ is the number of walkers at the time step $\tau$ and $\langle V(\tau) \rangle$ represents the average
potential energy of all of the walkers at that step.

In this study, the imaginary time step $\Delta\tau$ = 10 a.u. and $\alpha$ = 10. The random walkers are initiated at the Prism, Cage, and Book isomers respectively, and for each isomer, three DMC calculations are performed. In each DMC calculation, the number of walkers is 50,000, and these walkers are equilibrated for 10,000 time steps followed by 50,000 propagation steps. Previous studies show that this number of walkers may produce systematic uncertainties in terms of the absolute values of the ZPEs due to finite number of walkers; however, the relative ZPEs between different isomers are reliable, based on previous studies.\cite{Mallory2015, wanghex} 
In addition to the systematic uncertainties, there are statistical uncertainties because the reference energy oscillates around the ZPE. The statistical uncertainty is estimated as the standard deviation of the 3 DMC runs for the same isomer.

\section*{Details of the molecular dynamics simulations}
We interfaced the q-AQUA water potential with the i-PI software\cite{i-pi} to enable both classical and path integral molecular dynamics simulations to be performed for bulk water. Both classical and path integral molecular simulations were conducted with 256 water molecules in a periodically replicated simulation box with the experimental density set to be that at 300 K. For the classical MD simulation, at each temperature, we first propagated the trajectory for 100 ps to reach thermal equilibrium and then ran three independent trajectories for 100 ps each. The static and dynamical properties were calculated as an average over the three trajectories. The stochastic velocity rescaling thermostat is employed for the NVT classical MD simulations. For the path integral and also the ring polymer MD simulations (PIMD and RPMD), the trajectories were propagated in the NVT ensemble with a path-integral white noise Langevin thermostat (PILE-g), and 8 beads were included. For each temperature, we ran a single trajectory for 50 ps to achieve thermal equilibrium and then ran three independent trajectories for 50 ps with time step of 0.25 fs.

\subsection*{Mean square displacement and self-diffusion constants}
The self diffusion coefficient, $D$, of liquid water can be calculated from:
\begin{equation}
    D = \frac{1}{3}\int_0^{\infty} \langle \textbf{v}(0)\cdot \textbf{v}(t) \rangle dt = \frac{1}{6}\lim_{t \rightarrow \infty}\frac{d\langle \parallel \textbf{r}(t)-\textbf{r}(0)\parallel^2 \rangle}{dt}
\end{equation}
where $\langle \parallel \textbf{r}(t)-\textbf{r}(0)\parallel^2 \rangle$ is the mean square displacement (MSD). For each trajectory, we used the center of mass of each water molecule to calculate the MSDs and conducted linear fits to obtain the slope of the MSD curve. The self-diffusion constant D is simply $1/6$ of the MSD slope and the final values reported in the main text are from the averaged values over different trajectories. 

\begin{figure}[htbp!]
\begin{center}
\includegraphics[width=0.6\textwidth]{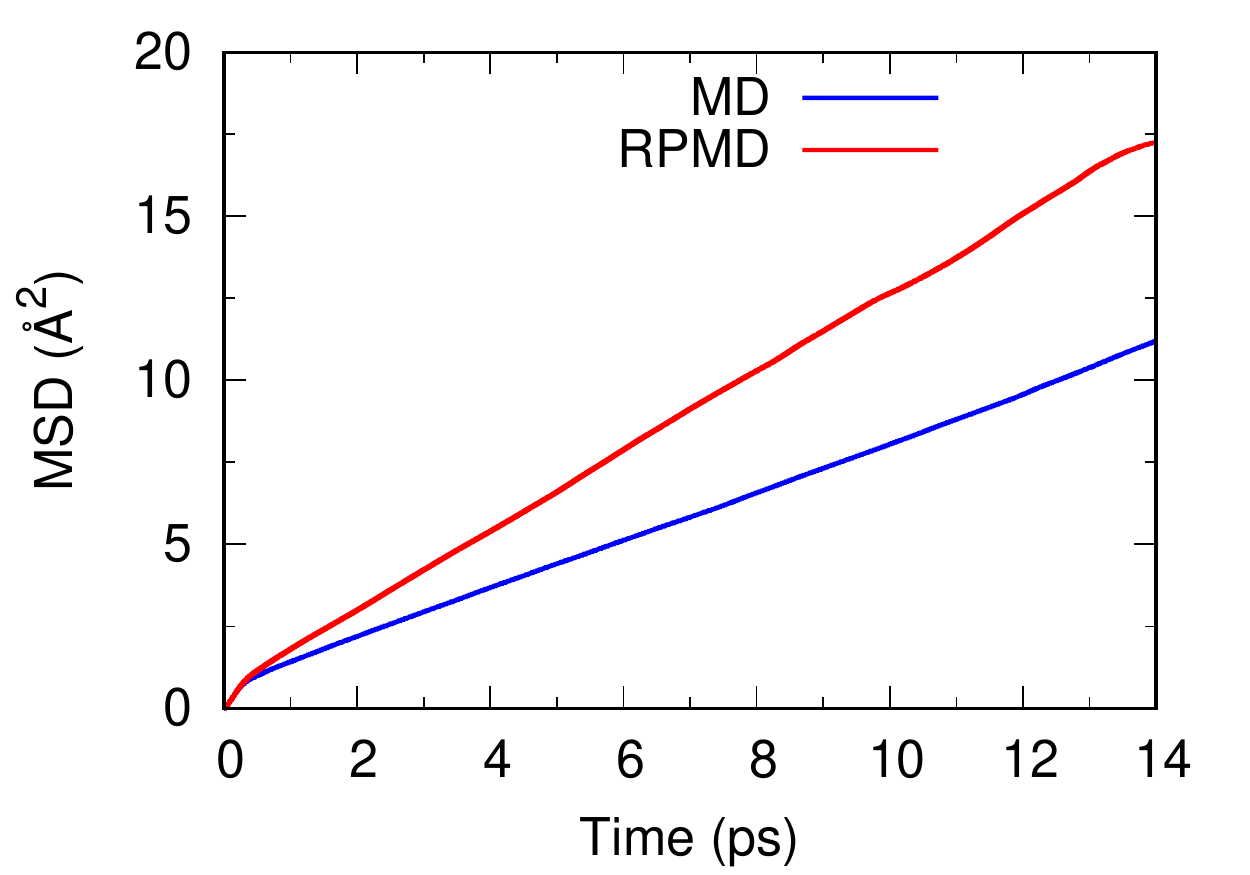}
\end{center}
\caption*{Figure S12: Mean square displacements for classical MD (blue) and PIMD (red) at the temperature of 298 K as a function of time.}
\end{figure}

Figure S12 shows two typical MSD curves from classical and ring polymer MD simulation at 298 K. As seen, both of them show linear behavior with the propagation of time and it is reliable to perform a linear fit for obtaining the diffusion coefficient. However, it should be noted that to obtain more accurate and converged diffusion constants of liquid water, longer time trajectories are required. We will conduct more extensive MD simulations in the future.

\subsection*{Second-order rotational time correlation functions and orientational relaxation time}

The second-order rotational time correlation function for OH bond is calculated as:
\begin{equation}
    C_2(t)=\big \langle P_2(\textbf{u}_{\text{OH}}(t)\cdot \textbf{u}_{\text{OH}}(0)) \big \rangle
\end{equation}
where $\textbf{u}_{\text{OH}}(t)$ is the unit vector along each OH bond, and $P_2$ is the second-order Legendre polynomial. 

\begin{figure}[htbp!]
\begin{center}
\includegraphics[width=0.7\textwidth]{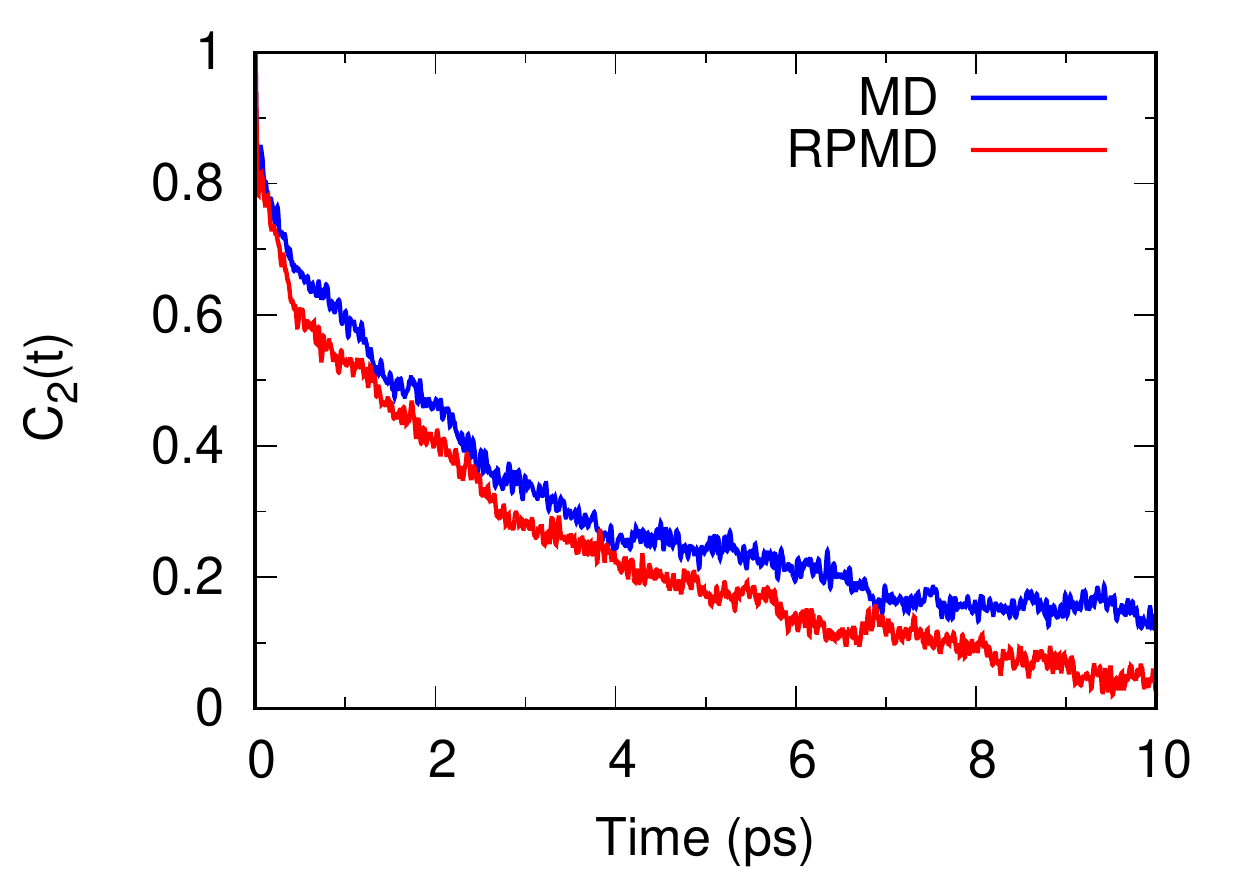}
\end{center}
\caption*{Figure S13: Typical second-order rotational time correlation functions, $C_2(t)$, from classical and ring polymer MD simulations at 298 K}
\end{figure}

Figure S13 shows two typical second-order rotational time correlation functions at 298 K from classical MD and RPMD simulations. To predict the orientational relaxation time, $\tau_2$, for both classical MD and RPMD simulations, we fit the correlation function $C_2(t)$ to a single exponential function first. The fitted exponential function was used to describe the long-time decay of $C_2(t)$ for times longer than 10 ps. We then numerically integrated the $C_2(t)$ function covering both short-time and long-time regions to obtain $\tau_2$.

\subsection*{Many body effects in liquid water}
As mentioned above, the q-AQUA water potential is constructed from 1-b, intrinsic 2-b, 3-b and 4-b interactions. It becomes natural to investigate how different components of potential influence the properties of liquid water. In Figure S14, we show both the OO radial distribution function and the MSD curves for liquid water at 298 K from classical MD simulation with different levels of potentials. As seen, using only 1-b + 2-b, both the static structural properties and dynamic properties of liquid water deviate considerably from the experiment. In particular, the peak positions of the OO radial distribution function do not agree with the experimental results. The slope of MSD curve is around 6 times of that when using full q-AQUA potential, where the latter is proven to have a diffusion coefficient close to the experiment. Thus, the use of only the 1-b + 2-b interactions cannot describe the important properties of liquid water -- higher body interactions are needed. This is verified by the fact that when  the 3-b interaction is added (see blue lines in Figure S14), both the radial distribution function and the MSD curve reach reasonable agreement with experiment. The self diffusion coefficients calculated from the 1-b + 2-b + 3-b interactions with classical MD simulations at different temperatures are listed in Table S4. It can be seen that when the PES is truncated at three-body level, it can already provide qualitatively reasonable diffusion coefficients comparing with the experiment. 

Finally, when the full q-AQUA potential is employed with the inclusion of 4-b interactions, Figure S14 shows that classical MD simulations give a more localized OO radial distribution function with higher distribution around 4-5 \AA. The slope of the MSD is lowered compared with that calculated with the 1-b + 2-b + 3-b interactions, and the diffusion constant at 298 K is also further decreased. These observations call into question of the role of the 4-b interaction and the adequacy of a classical MD simulation for accurately capturing all the static and dynamic properties. 

\begin{figure}[htbp!]
\begin{center}
\includegraphics[width=1.0\textwidth]{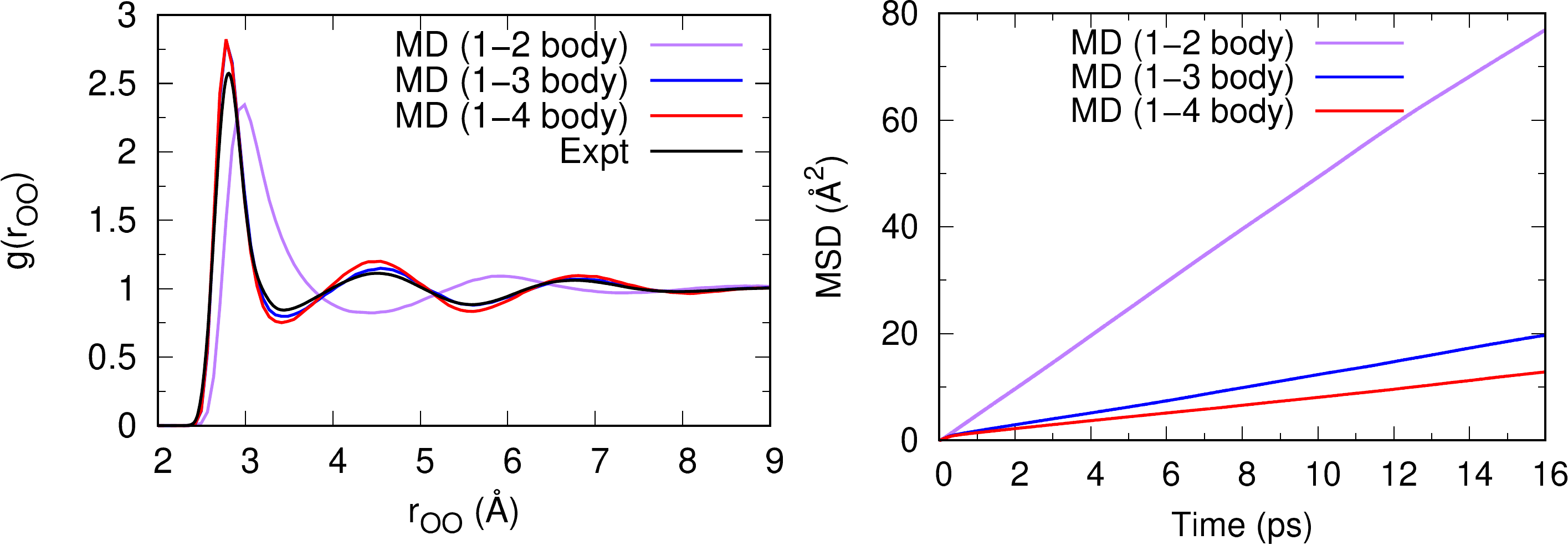}
\end{center}
\caption*{Figure S14: Effects of 2-b, 3-b, and 4-b interactions on the static structural and dynamical properties of liquid water at 298 K from classical MD simulations. Left: the OO radial distribution function (Experimental data are from Ref. \citenum{Skinner2013}), Right: the mean square displacement (MSD) as a function of time. MD (1-2 body) is the MD simulation using only 1-b + 2-b interactions. MD (1-3 body) is the MD simulation using  1-b + 2-b + 3-b interactions. MD (1-4 body) is the MD simulation using full q-AQUA potential with all 1-b, 2-b, 3-b and 4-b interactions included}
\end{figure}

\subsection*{Nuclear quantum effects in liquid water}
Figure S15 shows the effect of using 1- through 3-b interactions (left panel) or 1- through 4-b interactions (right panel) and using either classical molecular dynamics (MD) or quantum path-integral molecular dynamics (PIMD).  For both the 1- through 3-body and 1- through 4-body calculations, the quantum PIMD calculation is in substantially better agreement with experiment than that with classical MD.  

\begin{figure}[htbp!]
\begin{center}
\includegraphics[width=1.0\textwidth]{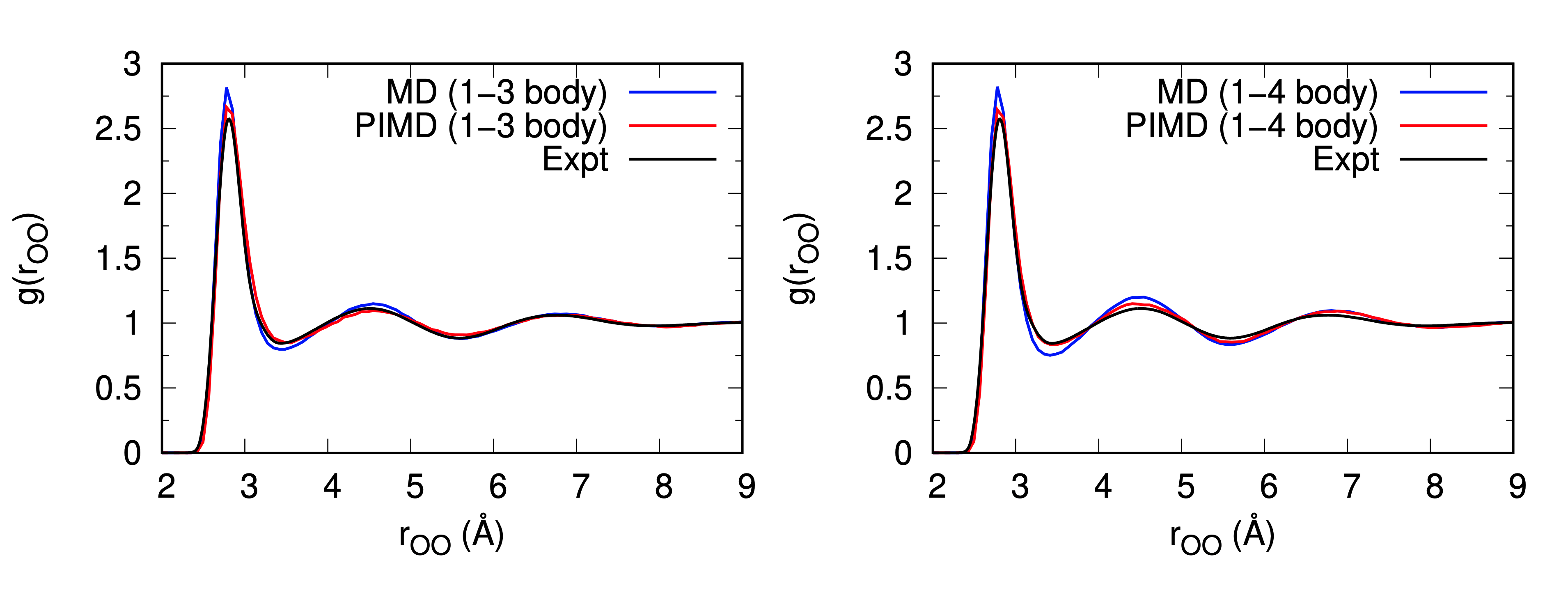}
\end{center}
\caption*{Figure S15: Nuclear quantum effects on the OO radial distribution function for  liquid water at 298 K from classical MD and PIMD simulations. Left: using potential with 1-b + 2-b + 3-b interactions. Right: using full q-AQUA potential with all 1-b, 2-b, 3-b and 4-b interactions included.
Experimental data are from Ref. \citenum{Skinner2013}}
\end{figure}

\begin{table}[htbp!]
\centering
\caption*{Table S4: Self-diffusion constant (\AA$^2$/ps) of liquid water at different temperatures using different components of the potential and different methods
\label{tab:diffusion}
}

\begin{threeparttable}
\begin{tabular*}{0.8\columnwidth}{@{\extracolsep{\fill}} c  c c c  }
\hline
\hline
 \multicolumn{4}{c}{Use full q-AQUA potential (1-b  + 2-b + 3-b + 4-b)}\\  
 \hline
Temperature (K) & Classical  & RPMD  &  Expt.\tnote{a}   \\
\hline
278 & 0.080 $\pm$ 0.016 & 0.130 $\pm$ 0.015 & 0.131 \\
288 & 0.102 $\pm$ 0.017 & 0.177 $\pm$ 0.015 & 0.177 \\
298 & 0.145 $\pm$ 0.012 & 0.226 $\pm$ 0.020 & 0.230\\
320 & 0.248 $\pm$ 0.011 & 0.331 $\pm$ 0.016 & 0.360 \\
\hline
\multicolumn{4}{c}{Use partial q-AQUA potential (1-b + 2-b + 3-b)}\\  
\hline
Temperature (K) & Classical  & RPMD  &  Expt.\tnote{a}   \\
\hline
278 & 0.088 $\pm$ 0.006 & 0.170 $\pm$ 0.014 & 0.131 \\
288 & 0.131 $\pm$ 0.011 & 0.216 $\pm$ 0.010 & 0.177 \\
298 & 0.188 $\pm$ 0.012 & 0.242 $\pm$ 0.012 & 0.230\\
320 & 0.332 $\pm$ 0.031 & 0.344 $\pm$ 0.020 & 0.360 \\
\hline
\hline
\end{tabular*}
\begin{tablenotes}
\item[a] from Ref. \citenum{Mills1973} and Ref. \citenum{Holz2000}
\end{tablenotes}

\end{threeparttable}
\end{table}

\subsection*{Computational cost of q-AQUA water potential}
Table S5 shows the computation cost of the q-AQUA model for both energy and gradient calculations for a 256-water system, using single core and multicores of 2.4 GHz Intel Xeon processor. Two calculations are shown, one with periodic boundary conditions (PBC) and one without.  The number of n-body configurations considered in the calculation is shown in the second column and is limited by a cut-off for the maximum OO distance.  For the 2-b interaction, in gas-phase cluster calculations there is no cut-off because the potential is switched in the long range to the dipole-dipole interaction.  In the MD simulation, a 14 \AA ~cut-off was used.  Figs. S3 and S4 show that the potential is essentially zero at this distance.  In the 3-b calculation, the cut-off limit was 9 \AA, at which distance the potential is also exceedingly small, as seen from Fig S8.  For the 4-b interaction, the switching range to zero for the MD simulations was set s 5.0--6.0 \AA~ in order to limit the number of calculations needed. A longer range of 5.8--7.5 \AA~ was used in for the gas-phase cluster work. In the MD simulations, the same cut-off maximum distances were used in calculations both with and without PBCs. Reverse derivative methods were used in all calculations with the exception of long-range 2-b ones that are based on the dipole-dipole interactions.

\begin{table}[htbp!]
\centering
\caption*{Table S5: The computation cost of the q-AQUA method for energy and gradient calculations of a 256 water system
}
\begin{threeparttable}
\begin{tabular*}{\columnwidth}{@{\extracolsep{\fill}} l c | c c |c  c }

\hline
\hline
  \multicolumn{2}{c|}{No PBC} &  \multicolumn{2}{c|}{Time for energy (s)}  & \multicolumn{2}{c}{Time for energy+gradient (s)}   \\
\hline
Component & Number  &  1 core  & 8 core  &   1 core  & 8 core  \\
1-b & 256 & 0.002 & 0.002\tnote{a} & 0.003 & 0.003\tnote{a}\\
2-b & 32640 & 0.23 & 0.02 & 0.72 & 0.08\\
3-b & 84051 & 0.42 & 0.05 & 1.94 & 0.26\\
4-b & 115922 & 1.26 & 0.17 & 4.34 & 0.52\\
\hline
\multicolumn{2}{c|}{Total}  & 2.00 & 0.35 & 7.12 & 0.97\\
\hline
\hline
  \multicolumn{2}{c|}{With PBC} &  \multicolumn{2}{c|}{Time for energy (s)}  & \multicolumn{2}{c}{Time for energy+gradient (s)}   \\
\hline
Component & Number  &  1 core  & 8 core  &   1 core  & 8 core  \\
1-b & 256 & 0.002 & 0.002\tnote{a} & 0.003 & 0.003\tnote{a}\\
2-b & 48919 & 0.38 & 0.04 & 1.18 & 0.18\\
3-b & 195058 & 0.84 & 0.12 & 5.30 & 0.70\\
4-b & 268304 & 2.91 & 0.37 & 11.55 & 1.31\\
\hline
\multicolumn{2}{c|}{Total}  & 4.52 & 0.83 & 18.05 & 2.49\\
\hline
\hline
\end{tabular*}
\begin{tablenotes}
\item[a] 1-b terms are not parallelized 
\end{tablenotes}
\end{threeparttable}
\end{table}

\clearpage
\bibliography{refs_SI}



